\documentclass{aa}  

\usepackage{graphicx}
\usepackage{txfonts}
\usepackage{multirow}
\usepackage{longtable,lscape}

\begin{document}

\title{NGC~3105: a young open cluster with low metallicity\thanks{This research is partially based on observations made with the New Technology Telescope under programme 090.D-0302(A) 
and the MPG/ESO 2.2-meter Telescope under programme 095.A-9020(A). Both telescopes are operated at La Silla Observatory (Chile) by the European Southern Observatory.}}

\author{
J. Alonso-Santiago\inst{1}
\and A. Marco\inst{1}
\and I. Negueruela\inst{1}
\and H. M. Tabernero\inst{1}
\and N. Castro\inst{2}
\and V. A. McBride\inst{3,4,5}
\and A. F. Rajoelimanana\inst{4,5,6}
}

\institute{
Dpto de F\'{i}sica, Ingenier\'{i}a de Sistemas y Teor\'{i}a de la Se\~{n}al, Escuela Polit\'ecnica Superior, Universidad de Alicante, Carretera de San Vicente del Raspeig s/n, 03690, Spain \\
\email{javier.alonso@ua.es} \\
\and Department of Astronomy, University of Michigan, 1085 S. University Avenue, Ann Arbor, MI 48109-1107, USA\\
\and Office of Astronomy for Development, International Astronomical Union, Cape Town, South Africa\\
\and South African Astronomical Observatory, PO Box 9, Observatory, 7935, South Africa\\
\and Department of Astronomy, University of Cape Town, Private Bag X3, Rondebosch, 7701, South Africa\\
\and Department of Physics, University of the Free State, PO Box 339, Bloemfontein 9300, South Africa
}

\date{}

 
  \abstract
   {NGC~3105 is a young open cluster hosting blue, yellow and red supergiants. This rare combination makes it an excellent laboratory to constrain evolutionary models of high-mass stars.
   It is poorly studied and fundamental parameters such as its age or distance are not well defined.}
   {We intend to characterize in an accurate way the cluster as well as its evolved stars, for which we derive for the first time atmospheric parameters and chemical abundances.}
   {We performed a complete analysis combining $UBVR$ photometry with spectroscopy. We obtained spectra with classification purposes for 14 blue stars and high-resolution spectroscopy for 
   an in-depth analysis of the six other evolved stars.}
   {We identify 126 B-type likely members within a radius of 2.7$\pm$0.6\,arcmin, which implies an initial mass, $M_{\textrm{cl}}\approx$4\,100\,M$_{\sun}$.
   We find a distance of 7.2$\pm$0.7\,kpc for NGC~3105, placing it at $R_{\textrm{GC}}$=10.0$\pm$1.2\,kpc. Isochrone fitting supports an age of 28$\pm$6\,Ma, implying masses around
   9.5\,M$_{\sun}$ for the supergiants. A high fraction of Be stars ($\approx$25$\%$) is found at the top of the main sequence down to spectral type b3.
   From the spectral analysis we estimate for the cluster an average $v_{\textrm{rad}}$=$+46.9\pm0.9$\,km\,s$^{-1}$ and a low metallicity, [Fe/H]=$-$0.29$\pm$0.22. We also have
   determined, for the first time, chemical abundances for Li, O, Na, Mg, Si, Ca, Ti, Ni, Rb, Y, and Ba for the evolved stars. The chemical composition of the cluster is consistent with that 
   of the Galactic thin disc. An overabundance of Ba is found, supporting the enhanced $s$-process.}
   {NGC~3105 has a low metallicity for its Galactocentric distance, comparable to typical LMC stars. It is a valuable spiral tracer in a very distant region of the 
   Carina-Sagittarius spiral arm, a poorly known part of the Galaxy. As one of the few Galactic clusters containing blue, yellow and red supergiants, it is massive enough 
   to serve as a testbed for theoretical evolutionary models close to the boundary between intermediate and high-mass stars.}


   \maketitle
%

\section{Introduction}\label{intro}

Red supergiants (RSGs) are believed to correspond to the He-core burning phase of stars with masses between $\approx8$\,--\,$25\:$M$_{\sun}$ \citep{lev05}. They represent a very important population,
because they are easy to identify and, being extremely bright in the infrared ($M_{K}$ up to $\sim-11$), can be seen through very heavy obscuration. Their numbers and properties are a fundamental
test to models of high-mass star evolution \citep{lev05,mey11}. Over the last years, several clusters of RSGs have been found in the inner Galaxy \citep{dav07,cla09a,alic7,alic10}. These Red Supergiant
Clusters (RSGCs) are concentrations of RSGs (between 8 and 26 such objects) in regions of very high obscuration. The high extinction has prevented so far the study of cluster members other
than the RSGs, rendering the determination of cluster parameters very insecure \citep{dav07}. Clusters with similar RSG populations and much lower obscuration are very rare, with VdBH\,222 \citep{vdBH222} the only example known till now. The lower obscuration allows the detection of the intrinsically blue population and thus 
much stronger constraints on the cluster mass.
 
The RSG phase is very short compared to the lifetime of the stars; RSGs are thus scarce. Most well-studied young open clusters in the 10\,--\,25~Ma age range have at most two (and in many cases no) RSGs \citep{egg02}. However, RSGCs contain so many because they are more massive than typical open clusters in the solar neighbourhood. Population synthesis simulations using a Kroupa IMF and sophisticated modelling predict that we will on average find $5\pm2$~RSGs in
a  10\,$000\:$M$_{\sun}$ cluster \citep{cla09b}. Based on this, Stephenson 2, with 26 RSGs, would have $M_{\textrm{cl}}\approx50\,000\:$M$_{\sun}$. Such values, however, are highly tentative, because the distances to the RSGCs have been calculated via indirect methods, such as radial velocities \citep{dav07} or average extinction \citep{alic7}. Such methods imply ages around 20~Ma for most of them. Observations reveal much random variation in the number of RSGs. A cluster as massive as h~Per ($M_{\textrm{cl}}\approx\,7\,000\:$M$_{\sun}$) contains no RSGs, even though there are 5 RSGs within the halo surrounding it and its twin $\chi$ Per, an area that may perhaps contain 20\,000\,M$_{\sun}$ \citep{sle02}. In contrast, the only moderate-extinction cluster 
known to contain 5 RSGs, NGC\,7419, has a similar mass $\approx7\,000\:$M$_{\sun}$ \citep{7419}. This cluster  presents some peculiarities that could also be due to statistical fluctuations:  it lacks blue supergiants (BSGs), 
while the fraction of BSGs to RGSs is $>1$ in the overall Milky Way cluster population \citep{egg02}; it also presents a very high fraction (the highest known) of Be stars (stars with emission lines). On the other hand, these characteristics, as well as
the ratio of RSGs to total mass, might vary from cluster to cluster according to unknown physical properties (binarity, average rotational velocity, etc). 

NGC\,3105 is a poorly-studied, faint and compact open cluster located in the constellation of Vela [$\alpha$(2000)\,=\,10h\:00m\:39s, 
 $\delta$(2000)\,=\,$-54\degr\:47\arcmin\:18\arcsec$; $\ell=279\fdg92$, $b=0\fdg26$]. \citet{Mo74} performed the first study of NGC~3105 employing photographic and photoelectric 
$UBV$ photometry. Later, \citet{Fi77} upgraded this study providing more photoelectric $UBV$ observations and slit spectroscopy for a few stars. Both works agree on the colour excess,
$E(B-V)\approx1.1$, but differ significantly when estimating the distance ($8.0\pm1.5$~kpc and $5.5\pm0.8$~kpc, respectively). In any case, the reddening to NGC~3105 is quite low for such high distances. 
Between the two studies, 34 stars were measured photoelectrically. \citet{Fi77} identified 15 likely blue members inside a cluster radius of $0\farcm9$ and seven other likely members at higher distances from the centre.
They also identified five candidate yellow and red supergiants and some emission-line star candidates. They give a spectral type at turn-off of B2, slightly later than 
their previous \citep{Mo74} estimate of b1\footnote{We follow the classical notation to distinguish spectral types from spectroscopy (capital letters) from those derived 
via photometry (lowercase letters).}, which explains the shorter distance.

Further work on NGC\,3105 is scarce. There are only two papers reporting modern CCD observations. \citet{Sa01} took $BVRI$ CCD 
photometry down to $V\approx18$~mag, finding 37 likely members. In combination with the $U$ values from \citet{Fi77}, and obtaining a similar reddening, they estimated an even longer distance of $9.5\pm1.5$~kpc. More recently,
\citet{Pau05} used CCD photometry in the $\Delta a$ photometric system to look for peculiar stars. They placed the cluster at $8.53\pm1.03$~kpc, estimating a slightly lower reddening, $E(B-V)$=0.95\,$\pm$\,0.02, from the 48 cluster members selected. 

The two CCD studies agree on an age around 20--25 Ma, which, despite this discrepancy in the distance determination, is compatible with the earliest spectral type 
on the main sequence (MS) around B2 found by \citet{Fi77}. At this age, the evolved stars should have masses around $12\:$M$_{\sun}$, comparable to the $\approx12$\,--\,$16\:$M$_{\sun}$ for the RSGs found in the RSGCs.

Recently, \citet{poster_3105} carried out spectroscopic observations of the brightest stars of the cluster, finding two blue supergiants. With this,
NGC~3105 acquires a very high intrinsic astrophysical interest: on the one hand, because of the co-existence of evolved high-mass stars in the blue, yellow and red supergiant phase.
This is a very rare occurrence, because of the short duration of these phases. Only the starburst cluster Westerlund 1 has a full set of evolved high-mass stars \citep{cla05}. However, at
an age around 5 Ma, Westerlund 1 probes the evolution of stars with $M_{*}\approx35$\,--\,$40\:$M$_{\sun}$, while NGC~3105 would sample much lower masses. On the 
other hand, NGC~3105 represents a good template for RSGCs. It is rather less reddened than NGC~7419 and, unlike the latter, hosts BSGs. In addition, it is a valuable spiral tracer in a distant and poorly known part of the Galaxy. 

In this work we want to improve the characterisation of the cluster, by determining an accurate distance as well as other parameters such as age, size or mass. Morever, as the main goal, we are going to obtain the stellar properties of the cool supergiants, 
both atmospheric parameters and chemical abundances, in order to perform the most complete study of the cluster so far. 
To carry out these tasks, we have the deepest $UBVR$ photometry to date and spectroscopy for all the bright stars in the field, including high-resolution spectra of the cool supergiants.


\section{Observations and data}

\begin{figure*} 
  \centering         
  \includegraphics[width=15cm]{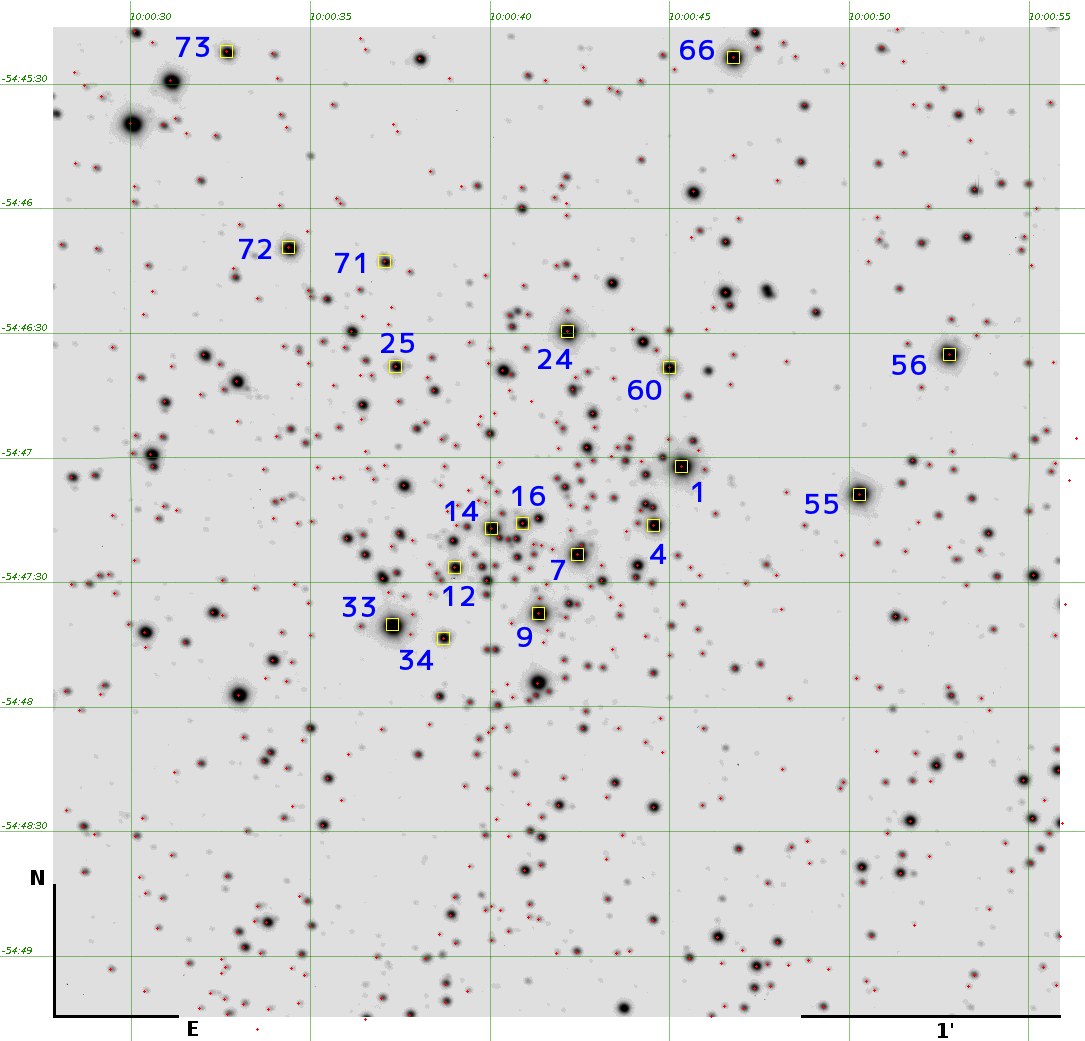}   
  \caption{Finding chart for the 607 stars with photometry in the field of 4.2$\arcmin$x4.2$\arcmin$ centred on NGC~3105 (red dots). Yellow squares highlight the stars for which we have obtained spectra, with the exception of
  stars 101 and 406, which fall outside the field of view covered by the photometry. The identification of each star corrresponds to the WEBDA numbering for this cluster. The
  image is one of our $V$-band frames. North is up and East is right.} 
  \label{3105}
\end{figure*}

\subsection{Optical photometry}
We obtained $UBVR$ photometry of NGC~3105 in visitor mode using the ESO Faint Object Spectrograph and Camera (EFOSC2) on the NTT 3.6-meter telescope at the La Silla Observatory 
(Chile) during three consecutive nights in 2013, from January 31 to February 02. EFOSC2 \citep{efosc} covers a field of view of $4\farcm1\times4\farcm1$ with a pixel scale of $0\farcs24$ pixel$^{-1}$.
Images of NGC~3105 were taken in several series of different exposure times in each filter, listed in Table~\ref{log}, to achieve accurate photometry for a broad magnitude 
range. Two standard fields from the list of \citet{landolt}, PG\,1047+003 and SA\,101, were observed to trace extinction and provide standard stars for the transformation. 
All these images were processed for bias and flat-fielding corrections with the standard procedures using the {\scshape ccdproc} package in {\scshape iraf}.\footnote{{\scshape iraf} 
is distributed by the National Optical Astronomy Observatories, which are operated by the Association of Universities for Research in Astronomy, Inc., under cooperative agreement
with the National Science Foundation.}. 

\begin{table}
\caption{Log of the photometric observations for NGC~3105 taken at the NTT.}
\begin{center}
\begin{tabular}{lccc}   
\hline\hline
\multirow{2}{*}{NGC~3105} & \multicolumn{3}{c}{RA\,=\,10$^h$00$^m$39$^s$  DEC\,=\,$-54\degr47\arcmin18\arcsec$}\\
 & \multicolumn{3}{c}{(J2000) \,\,\,\,\,\,\,\,\,\,\,\,\,\,\,\,\,\,\,\,\,\,\,\,\,\,\, (J2000)}\\
\hline
\multirow{2}{*}{Filter} & \multicolumn{3}{c}{Exposure times (s)}\\
 & Short & Intermediate  & Long \\
\hline
$U$ & 60 & 500 & 1800, 2400\\
$B$ & 12, 24 & 100& 400, 600, 800\\
$V$ &  3, 4, 6 & 18 &  90, 100\\
$R$ &  1, 3 & 9 &  12, 15\\
\hline
\end{tabular}

\label{log}
\end{center}
\end{table}

Aperture photometry using the {\scshape phot} package inside {\scshape daophot} was carried out on the standard stars with the same aperture (35 pixels) for each filter. For stars in the field of NGC~3105 we performed photometry by point-spread function (PSF) fitting using the {\scshape daophot} package \citep{daophot} provided by {\scshape iraf}. We used six-pixel apertures (similar to the full width at half maximum) for each image in all filters. To construct the PSF empirically, we selected bright stars (around 25) and discarded those with close companions or any other impediment. With the good PSF stars ($\approx$\,20), we determined an initial PSF by fitting the best function among the five options offered by the {\scshape psf} routine inside {\scshape daophot}. We allowed the PSF to be
variable (of order 2) across the frame because there may be systematic patterns of PSF variability with position on the chip. We then performed aperture correction for each frame in all filters.

Finally, we obtained the instrumental magnitudes for all stars. Using the standard stars in the Landolt fields, we carried out the atmospheric extinction correction and transformed the instrumental 
magnitudes (denoted with the subscript $i$) to the standard system with the {\scshape photcal} package inside {\scshape iraf}. The equations used are: 

\begin{multline}
U_i=(U-B)+(B-V)+V+(1.30\pm0.03)+ \\
     +(0.48\pm0.02)\,\cdot\,X_U-(0.046\pm0.007)\,\cdot\,(U-B) 
\end{multline}

\begin{multline}
B_i=(B-V)+V-(0.727\pm0.016)+ \\
+(0.233\pm0.012)\,\cdot\,X_B-(0.038\pm0.003)\,\cdot\,(B-V)
\end{multline}

\begin{multline}
V_i=V-(0.83\pm0.02)+(0.114\pm0.016)\,\cdot\,X_V-\\
      -(0.069\pm0.014)\,\cdot\,(B-V)
\end{multline}

\begin{multline}
R_i=V-(V-R)-(0.95\pm0.02)+  \\ 
+(0.07\pm0.02)\,\cdot\,X_R-(0.016\pm0.014)\,\cdot\,(V-R)
\end{multline}

In total, we have photometry for 607 stars, which are displayed in Fig.~\ref{3105} as red dots. In Table \ref{phot} we list the identification number of each star ($ID$), the
equatorial coordinates (RA, DEC), the values of $V$, $(B-V)$, $(U-B)$, and $(V-R)$, together with their uncertainties, and the number of measurements ($N$). The magnitudes and colours displayed in the table are the
weighted arithmetic means (using the variances as weights) of all individual measurement. The errors are expressed in terms of the standard deviation around this mean. 

When possible, we also show their 2MASS counterpart. The astrometric transformation, from pixel ($X,Y$) to celestial coordinates (RA, DEC), was done in a first step using the 
{\scshape xy2sky} task provided by {\scshape wcstools} \citep{wcs} and, in a second pass, with a subsequent correction employing as an astrometric reference 35 stars from the 2MASS catalogue, 
homogeneously distributed throughout the cluster field.
The typical error of the astrometric fit, in both coordinates, is around $0\farcs3$.

As mentioned in Sect.~\ref{intro} there are a few previous photometric studies for this cluster. We used the WEBDA database\footnote{Available at
\url{http://univie.ac.at/webda/}} \citep{Webda}, to download their data and compare their photometric values to ours. Moreover, we also included data from \citet{Ly73} who obtained photoelectric magnitudes of stars in the field around the cluster (but not the cluster  itself). In Table~\ref{dif_fotom} we list the differences (present work minus literature data) and the number of stars in common ($N$) with each work. We find no systematic deviations with respect to the photoelectric photometry, and a remarkable agreement with the values by \citet{Fi77}.

\begin{table*}
\caption{Comparison of different photometries for NGC~3105 (this work\,$-$\,literature), with $N$ being the number of stars in common.}
\begin{center}
\begin{tabular}{lccccccccc}   
\hline\hline
Reference & Phot. & $\Delta$\,$V$ & $N$ & $\Delta$\,$(B-V)$ & $N$ & $\Delta$\,$(U-B)$ & $N$ & $\Delta$\,$(V-R)$ & $N$\\
\hline
\citet{Ly73} & Pe & $0.02\,\pm\,0.04$  & 9  & $0.01\,\pm\,0.07$ & 10 & $-0.07\,\pm\,0.07$ & 8  & ---                  & ---\\
\citet{Mo74} & Pe & $0.03\,\pm\,0.10$  & 12 & $-0.04\,\pm\,0.04$ & 12 & $-0.12\,\pm\,0.22$ & 9  & ---                  & ---\\
\citet{Fi77} & Pe & $-0.00\,\pm\,0.10$ & 27 & $-0.05\,\pm\,0.06$ & 27 & $-0.01\,\pm\,0.14$ & 27 & ---                  & ---\\
\citet{Sa01} & CCD& $-0.13\,\pm\,0.14$ & 69 & $-0.07\,\pm\,0.11$ & 60 & ---                  & ---& $-0.14\,\pm\,0.09$ & 66\\
\hline
\end{tabular}

\label{dif_fotom}
\end{center}
\end{table*}

Following the method of \citet{Maria}, we estimate for our photometry a completeness around 90 per cent at these limiting magnitudes for each band: $U$=21.3, $B$=20.7, $V$=19.3 and $R$=18.7.

\subsection{2MASS data}

We completed our dataset with $JHK_{\textrm{S}}$ magnitudes from the 2MASS catalogue \citep{2MASS}. 
We cross-correlated the 2MASS catalogue with our photometric list of targets to find their infrared magnitudes. Only stars with good-quality photometry (i.e. without any ``$U$'' photometric flag in the catalogue) were selected.

\subsection{Spectroscopy}\label{espectros}

We took spectra for 20 stars out of which 18, covered by the photometry, are indicated by yellow squares in Fig.~\ref{3105}) by using three different spectrographs with several setups. These objects were selected based on the literature and location on the 2MASS colour-magnitude diagram (CMD)
in order to observe the brightest blue members as well as those evolved stars that have left the main sequence.

We collected spectra with the Cassegrain Spectrograph, mounted on the 1.9-meter telescope at the South African Astronomical Observatory (SAAO). The spectra were taken in visitor mode from April 11 to 18, 2012. The spectral range as well as the resolution of spectra depends on the diffraction grating used. We employed the gratings 7, 8 and 10 in order to 
observe the blue spectral region, the H$\alpha$ line and the Ca\,{\scshape{ii}} triplet, respectively. Table~\ref{setup} displays their characteristics. For wavelength calibration, we used CuAr and CuNe arc spectra interspersed between observations.  The reduction was carried out with standard {\scshape iraf} tools. 

EFOSC2 is a versatile instrument that also works as a long-slit spectrograph. We used grism 14 and the volume phase holographic (VPH) grism 19 to observe the blue spectral region with the highest resolution available. The grisms were used in combination with an off-centre slit to displace their nominal wavelength range to those shown in Table~\ref{setup}. 
The typical signal-to-noise ratios ($S/N$) obtained for these spectra are listed in Table~\ref{setup}. The EFOSC2 spectra were reduced following standard procedures with the {\scshape starlink} software
packages {\scshape ccdpack} \citep{draper} and {\scshape figaro} \citep{shortridge}. In addition, we obtained slitless spectroscopy of the field by combining the low-resolution grism 13 with the Gunn $r$-band filter. Two slitless images at different orientations were 
taken to minimise source overlapping.

Finally, we obtained high-resolution $\acute{e}chelle$ spectra using FEROS (Fiber Extended Range Optical Spectrograph) mounted on the ESO/MPG 2.2-meter telescope at the La Silla Observatory, 
in Chile. FEROS \citep{feros} covers the wavelength range from 3500 to 9200\,\AA{}, providing a resolving power of $R=48\,000$. This spectral region is covered in 39 orders, with 
small gaps between the orders appearing only at the longest wavelengths. The spectra were taken during two consecutive nights in May 29\,--\,30, 2015. 
The exposure times ranged from 2\,700 to 5\,400 seconds to achieve a typical $S/N\approx70$\,--\,80. 
The spectroscopic reduction was performed using the {\scshape feros-drs} pipeline based on {\scshape midas} routines comprising the usual steps of: bad pixel and cosmic
correction, bias and dark current subtraction, removal of scattered light, optimum order extraction, flatfielding, wavelength calibration using Th-Ar lamps exposures, 
rectification and merging of the $\acute{e}chelle$ orders.

\begin{table}
\caption{Instrumental setups for the different spectrographs used.}
\begin{center}
\begin{tabular}{lccc}   
\hline\hline
\multirow{2}{*}{Setup} & Range   & Dispersion          & \multirow{2}{*}{$S/N$}\\
                       & (\AA{}) & (\AA{}\,pix$^{-1}$) &    \\
\hline
\multicolumn{3}{c}{EFOSC2}\\
\hline
Grism 13   & 3685 -- 9315  & 2.77 & ---\\
Grism 14   & 3750 -- 5700  & 0.93 & 75\\
Grism 19   & 4235 -- 4875  & 0.34 & 100\\
\hline
\multicolumn{3}{c}{Cassegrain Spectrograph}\\
\hline
Grating 7  & 3093 -- 7105 & 2.72 & 110 \\
Grating 8  & 6425 -- 9279 & 2.21 & 80  \\
Grating 10 & 8306 -- 8927 & 0.47 & 40  \\
\hline

\end{tabular}

\label{setup}
\end{center}
\end{table}


\begin{table*}
\caption{Spectral types, spectrographs used and estimates of cluster membership are shown for stars observed spectroscopically in the field of NGC~3105. For a clear identification of 
every star both numbering, WEBDA and ours (ID), are shown.}
\begin{center}
\begin{tabular}{lccccccccc}   
  \hline\hline
    \noalign{\smallskip}
\multicolumn{2}{c}{Star} & \multirow{2}{*}{Spectral type} & \multicolumn{2}{c}{EFOSC2} & \multicolumn{3}{c}{SAAO} & \multirow{2}{*}{FEROS} & \multirow{2}{*}{Membership}\\
WEBDA   & ID & & Gr. 14 & Gr. 19 & Gr. 7 & Gr. 8  & Gr. 10 & \\
  \noalign{\smallskip}
  \hline
    \noalign{\smallskip}
1   & 431 & A0\,Ib     & \textbullet   & \textbullet   & \textbullet  & \textbullet   &              &    & y\\
4   & 418 & B2\,IVe        & \textbullet   & \textbullet   &              &               &              &   & y \\
7   & 358 & B1.5\,Ve       & \textbullet   & \textbullet   & \textbullet  &               & \textbullet  &    & y\\       
9   & 316 & K3\,Ib         &               &               &              &               & \textbullet  & \textbullet & y\\
12  & 229 & B2\,V          & \textbullet   &               &              &               &              &   & y \\
14  & 267 & B3\,III        & \textbullet   & \textbullet   & \textbullet  &  \textbullet  &              &    & y\\
16  & 298 & B2.5\,V        & \textbullet   &               &              &               &              &   & y \\
24  & 348 & K3\,Ib + B2\,V:& \textbullet   &               &              &               & \textbullet  & \textbullet & y\\
25  & 191 & B2\,V          & \textbullet   &               &              &               &              &    & y\\
33  & --- & M1\,Iab        &               &               &              &               & \textbullet  &   & y \\
34  & 221 & B2.5\,V        & \textbullet   &               &              &               &              &    & y\\
55  & 503 & F9\,Ib         &               &               &              &               & \textbullet  & \textbullet & y\\
56  & 548 & A0\,Ib     & \textbullet   & \textbullet   & \textbullet  &               & \textbullet  &    & y\\
60  & 426 & B2\,IVe?       &               & \textbullet   &              &               &              &   & y \\
66  & 459 & K5\,Ib         &               &               &              &               & \textbullet  & \textbullet& y\\          
71  & 184 & B2\,IV         & \textbullet   &               &              &               &              &   & y \\
72  & 121 & B1.5\,Ve       &               & \textbullet   & \textbullet  &               &              &    & y\\
73  & 181 & M2\,III        &               &               &              & \textbullet   & \textbullet  & \textbullet& n\\ 
101 & --- & B2\,Ve         & \textbullet   &               &              &               &              &    & y?\\
406 & --- & M6\,Ib     &               &               &              &               &              & \textbullet& y?\\
  \noalign{\smallskip}
\hline
\end{tabular}

\label{SpT}
\end{center}
\end{table*}

\section{Results}

Throughout this paper we use the WEBDA numbering to identify cluster stars observed spectroscopically. The designation in our photometry (ID) can be found in Tables~\ref{phot} and~\ref{phot_spt}. Table~\ref{SpT} lists both designations for stars with spectra.

\subsection{Spectral classification}

We obtained spectra for the brightest stars in the cluster. In total, we collected 41 spectra for 20 different stars (see Table~\ref{SpT} for spectrograph setups and spectral 
types assigned). Taking into account all the configurations used, we covered, at different resolution, a spectral range between 3100--9200\,\AA{} that includes the main spectral classification regions
for both blue and red stars. We estimate for our classification a typical uncertainty around one spectral subtype.

\subsubsection{Blue stars}

We took spectra of the blue stars in the field of NGC~3105 to study the upper main sequence and find the spectral type at which the main sequence turnoff (MSTO) point is
seen on the CMD.
We followed classical criteria of classification in the optical
wavelength range (4\,000\,--\,5\,000\,\AA), following \citet{Ja87} and \citet{gray}. From 13 low- and intermediate-resolution spectra taken with EFOSC2 (Gr14 and Gr19) and the SAAO Cassegrain Spectrograph 
(Gr7) we found two A supergiants (A0\,Ib) and 11 B-type stars.

With respect to the A supergiants, around spectral type A0 the \ion{Ca}{ii}~K-line is a notable feature while the profile of the Balmer lines is the primary luminosity 
criterion: the narrower their wings are, the higher the luminosity. The \ion{Fe}{ii} line at 4233\,\AA{}, the blends of \ion{Fe}{ii} and \ion{Ti}{ii} at 4172\,--\,8\,\AA{} and the 
\ion{Si}{ii}\,$\lambda$4128\,--\,30 doublet are enhanced in the supergiants, but at our resolution these features are not clearly separated, although visible in the spectra. 

\begin{figure*}  
  \centering         
  \includegraphics[width=15cm]{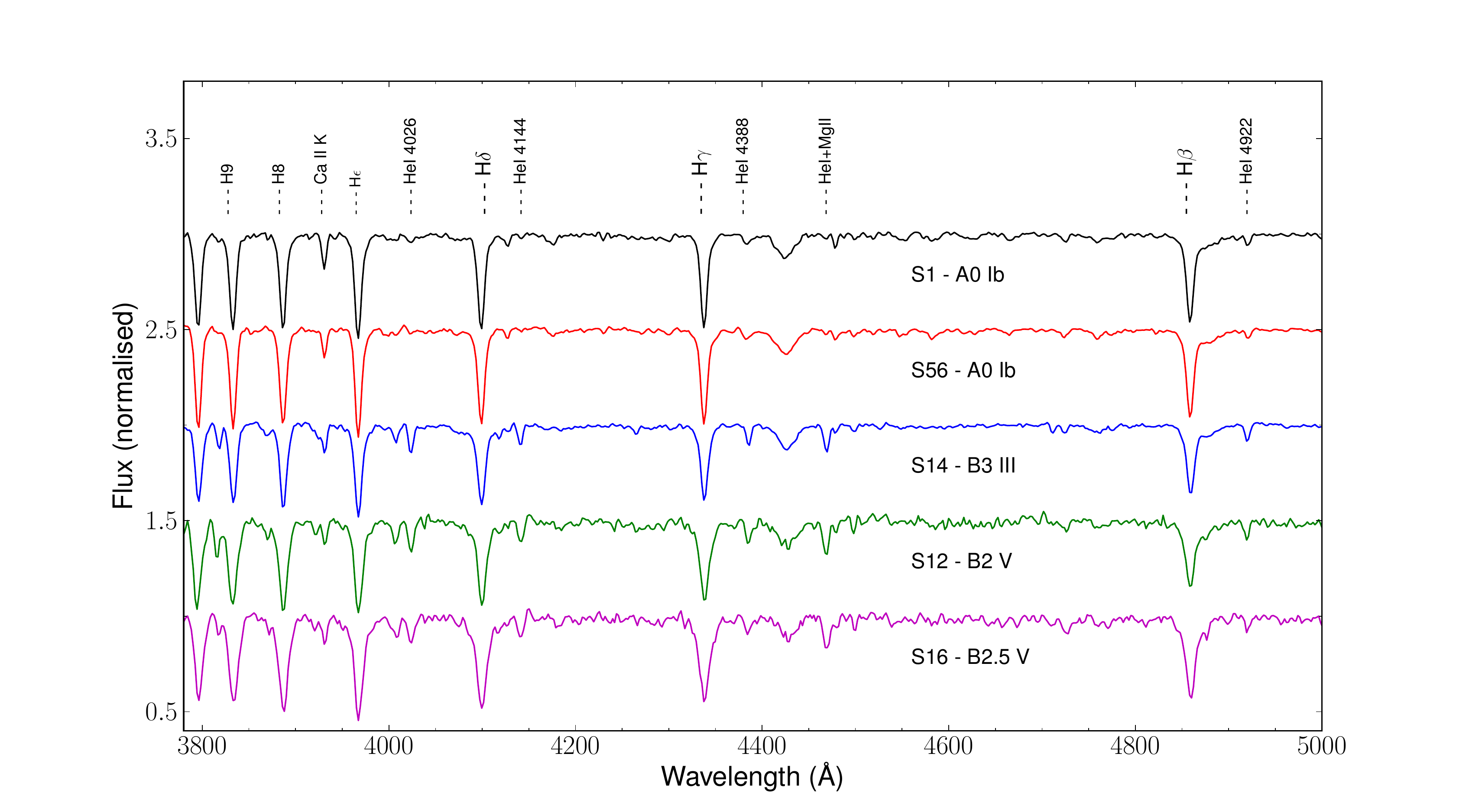}   
  \caption{Spectra of the brightest blue stars in the classical spectral classification region. These spectra have been obtained with grism 14 on EFOSC2. The most important lines are marked.}
  \label{esp_azules}  
\end{figure*}

Regarding the B-type stars, as we move toward later spectral types the ratio \ion{Mg}{ii}\,$\lambda$4481/\ion{He}{i}\,$\lambda$4471 and the strength of the \ion{Ca}{ii}~K-line increase, as \ion{He}{i} lines weaken. 
Our sample consists of early B stars with spectral types between B1.5 and B3. The line ratios used in this range to assign spectral types are \ion{Si}{ii}\,$\lambda$4128\,--\,30/\ion{Si}{iii}\,$\lambda$4553, 
\ion{Si}{ii}\,$\lambda$4128\,--\,30/\ion{He}{i}\,$\lambda$4121, \ion{N}{ii}\,$\lambda$3995/\ion{He}{i}\,$\lambda$4009, and \ion{He}{i}\,$\lambda$4121/\ion{He}{i}\,$\lambda$4144 \citep{Wa90}.
A sample of the brightest blue stars, including the A0 supergiants, is shown in Fig.~\ref{esp_azules}.

\begin{figure}  
  \centering         
  \includegraphics[width=9cm]{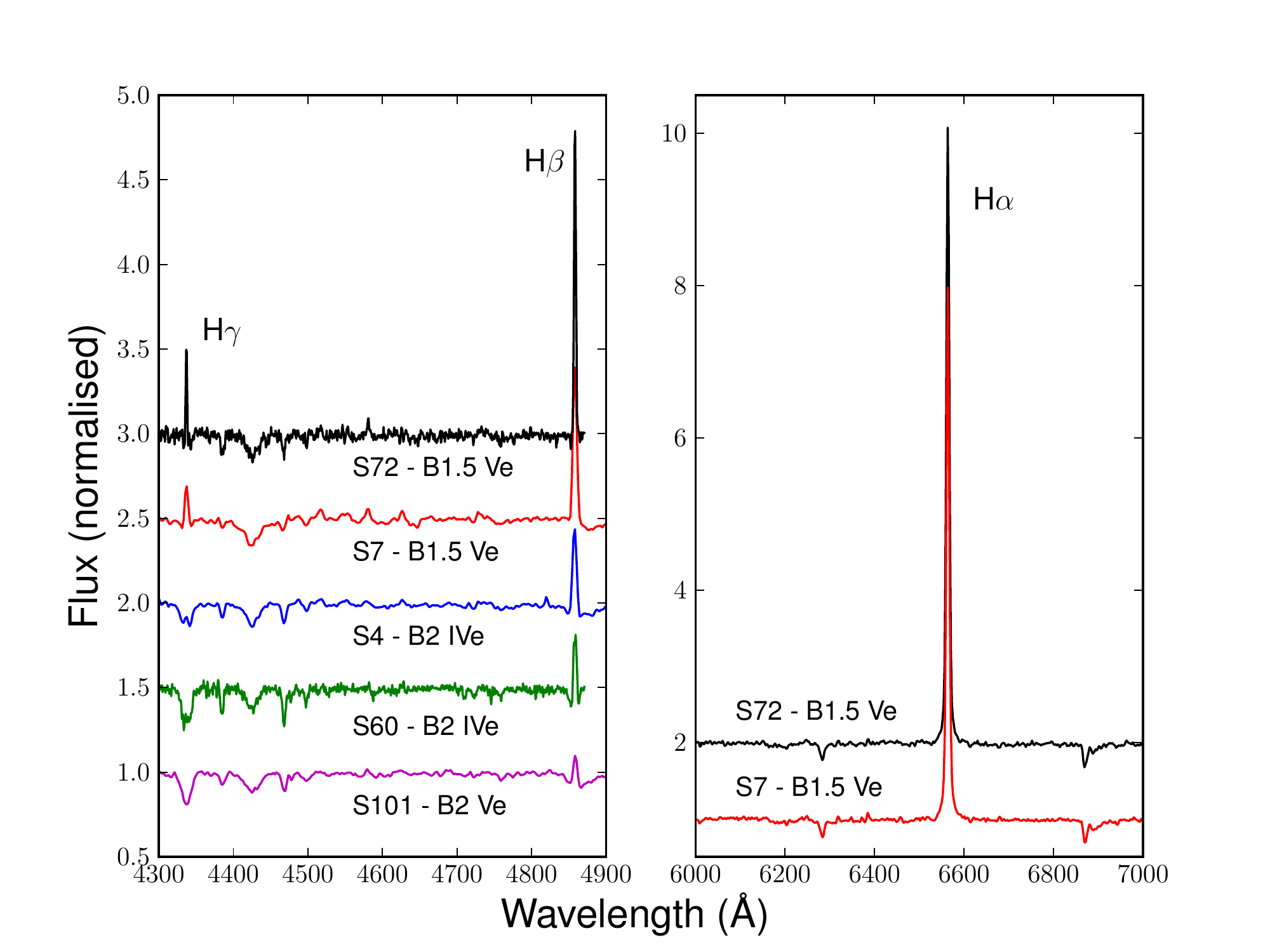}   
  \caption{Spectra of the Be stars in the field of NGC~3105. \textbf{Left:} EFOSC2 spectra around H$\beta$ and H$\gamma$ taken with grisms 14 (for stars 4, 7 and 101) and 19 (60 and 72). \textbf{Right:} Strong H$\alpha$ emission in stars 7 and 72. These spectra were taken with grating 7 on the SAAO Cassegrain Spectrograph.} 
  \label{be}  
\end{figure}

The earliest spectral types for MS stars, with no emission lines, were found around B2\,--\,B2.5. In addition, five of the B-type stars observed -- namely 4, 7, 60, 72, and 101 -- are emission-line stars (Be) with spectral types B1.5\,--\,B2. 
Stars 72 and 7 stand out because of their strong emission, showing $EW$(H$\alpha$) of $-73$ and $-67$\,\AA{} respectively. These are extreme values for classical Be stars. Only two objects among more than 50 Be stars monitored by \citet{barnsley13} have ever shown comparable values. In addition, six other Be stars have been identified in the slitless fields, 
namely stars 40, 78, 84, 92, 213 (this latter without a 2MASS counterpart) and 389. Among all these Be stars, that numbered as 101 is the only one far away from the cluster core; for this reason its membership is doubtful.
However, early-type stars like this are not abundant in the sky and, in addition, it presents a spectral type similar to that of the other brightest blue members. This fact might suggest some sort of 
relationship with the cluster despites its distance to the core. In Fig.~\ref{be} we show H$\alpha$, H$\beta$ and H$\gamma$ line profiles for all the Be stars with long-slit spectra. 

\subsubsection{Red stars}

We obtained spectra for seven evolved stars in the cluster field. We classified them by comparison with high-resolution standards from the atlas of \citet{munari} focusing on the near-infrared wavelength range, around the \ion{Ca}{ii} triplet (8480\,--\,8750\,\AA{}).
For this, we employed the spectra taken with the SAAO Cassegrain Spectrograph (Gr10) and FEROS, after degrading the latter to the resolution of the standards ($R\approx20\,000$). The triplet weakens toward later spectral types and lower luminosity classes \citep{Ja87}, but many other 
classification criteria are available in this range. Many features of \ion{Fe}{i} (i.e. lines at 8514, 8621 and 8688\,\AA{}) and \ion{Ti}{i} (8518\,\AA{}) become stronger with later spectral types \citep{carquillat}. In addition, we 
find very useful as classification criteria the ratios \ion{Ti}{i}\,$\lambda$8518\,/\,\ion{Fe}{i}\,$\lambda$8514 and \ion{Ti}{i}\,$\lambda$8734\,/\,\ion{Mg}{i}\,$\lambda$8736, which become larger with 
increasing spectral type.

\begin{figure*}  
  \centering         
  \includegraphics[width=15cm]{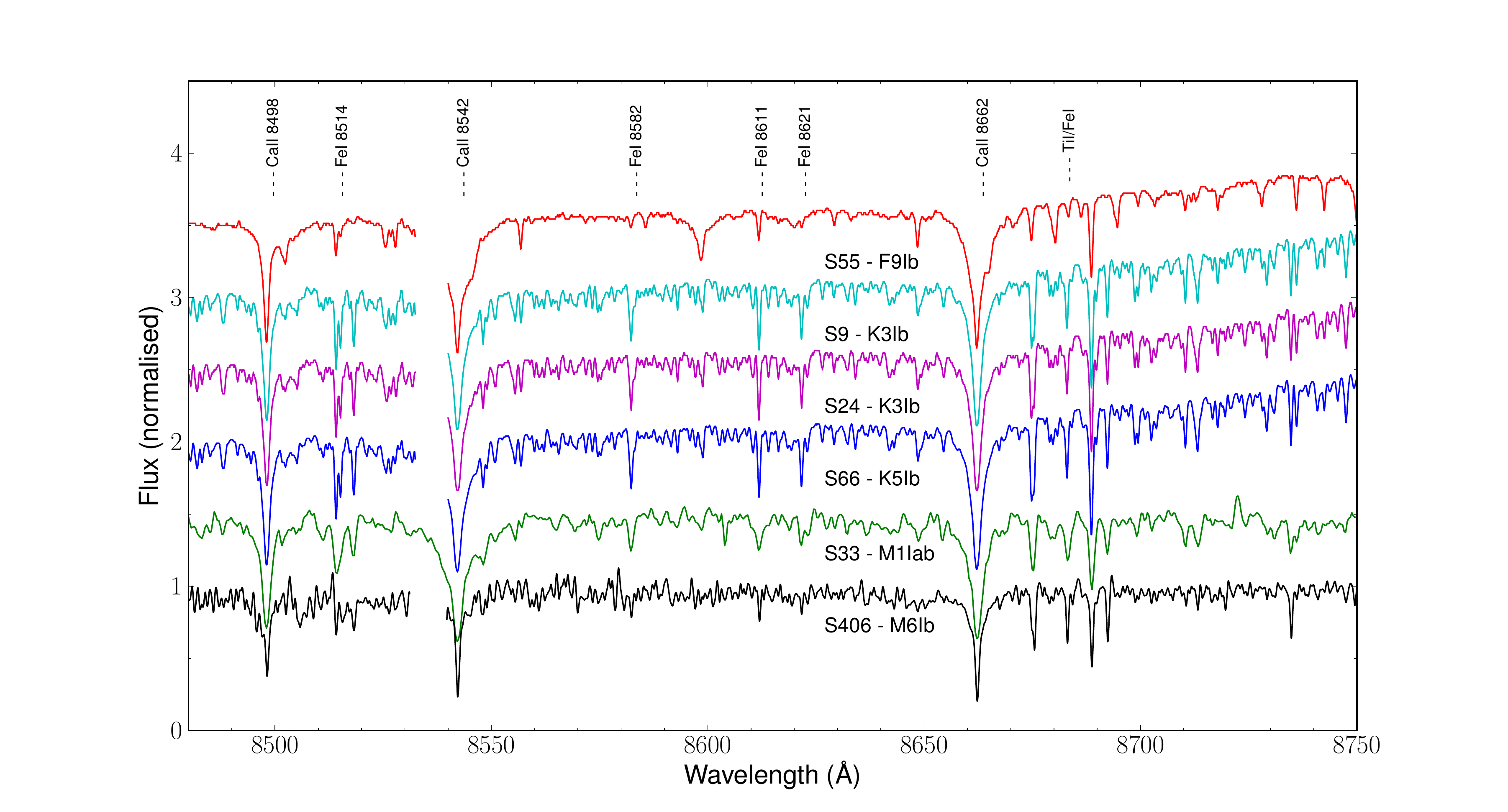}   
  \caption{Spectra of cool supergiant members around the \ion{Ca}{ii} triplet. All the spectra were taken with FEROS (note the gap between orders 37 and 38 around 8535\,\AA), except that of star 33, which was taken with grating 10 on the SAAO Cassegrain Spectrograph.
  \label{esp_rojos}}  
\end{figure*}

We found one yellow supergiant (star 55\,--\,F8\,Ib), three K-supergiants (9, 24 and 66) and three M-(super)giants (33, 73 and 406).
Star 24 presents a composite spectrum, as first noted by \citet{Fi77}, consisting of an early B star and a cool supergiant. The hot companion is only detected by the presence of the Balmer lines in the blue region, 
since the rest of the spectrum is dominated by the K-supergiant. The B-type component cannot thus be classified with accuracy. 
Figure~\ref{esp_rojos} displays the spectra of evolved members belonging to the cluster. Star 73 is not included since its radial velocity (RV) is not compatible with 
that of the cluster, as discussed later in Sect.~\ref{sec:rv}.

\subsection{Observational HR diagram: Cluster reddening, distance and age}\label{general}

\begin{figure}  
  \centering         
  \includegraphics[width=\columnwidth]{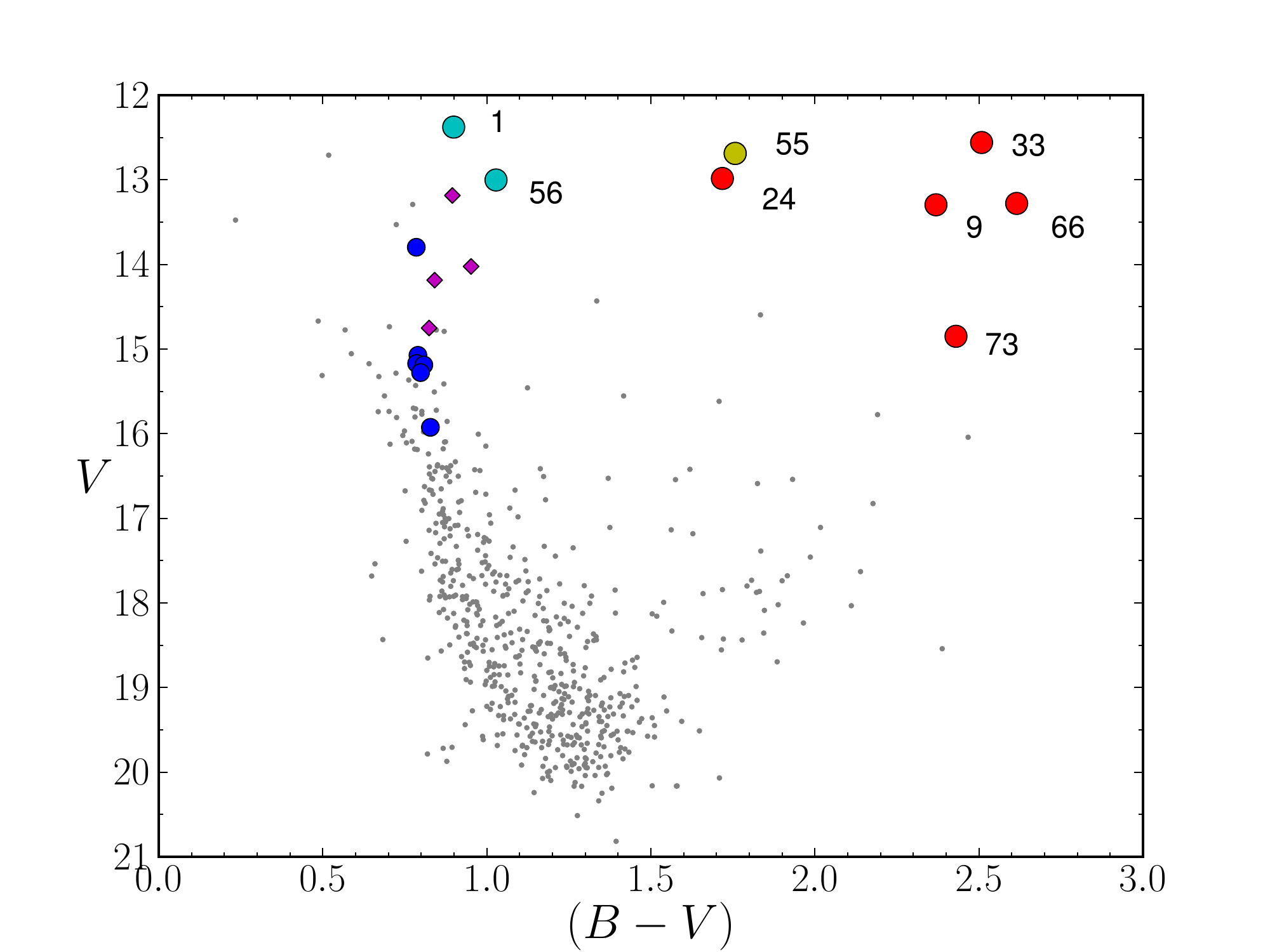}   
  \caption{$V/(B-V)$ diagram for all stars in the field of NGC~3105. Photometric data appear as gray dots. Stars for which we also have spectra are represented by small blue circles
  (upper main sequence B-type stars), magenta diamonds (Be stars) and coloured big circles (supergiant stars, whose colour depends on their spectral type: cyan for BSGs, yellow for the YSG,  
  and red for the RSGs).} 
  \label{CMD_opt}  
\end{figure}

We start the photometric analysis by plotting the $V/(B-V)$ colour-magnitude diagram (CMD) for all the stars in the field (see Fig.~\ref{CMD_opt}). We have highlighted stars spectroscopically observed, which are clearly following a sequence.
Given the spectral types of the earliest stars observed and the completeness levels calculated above, we estimate our photometry to be complete down to the typical magnitude of $\approx$\,A3\,V stars.

\begin{figure}  
  \centering         
  \includegraphics[width=\columnwidth]{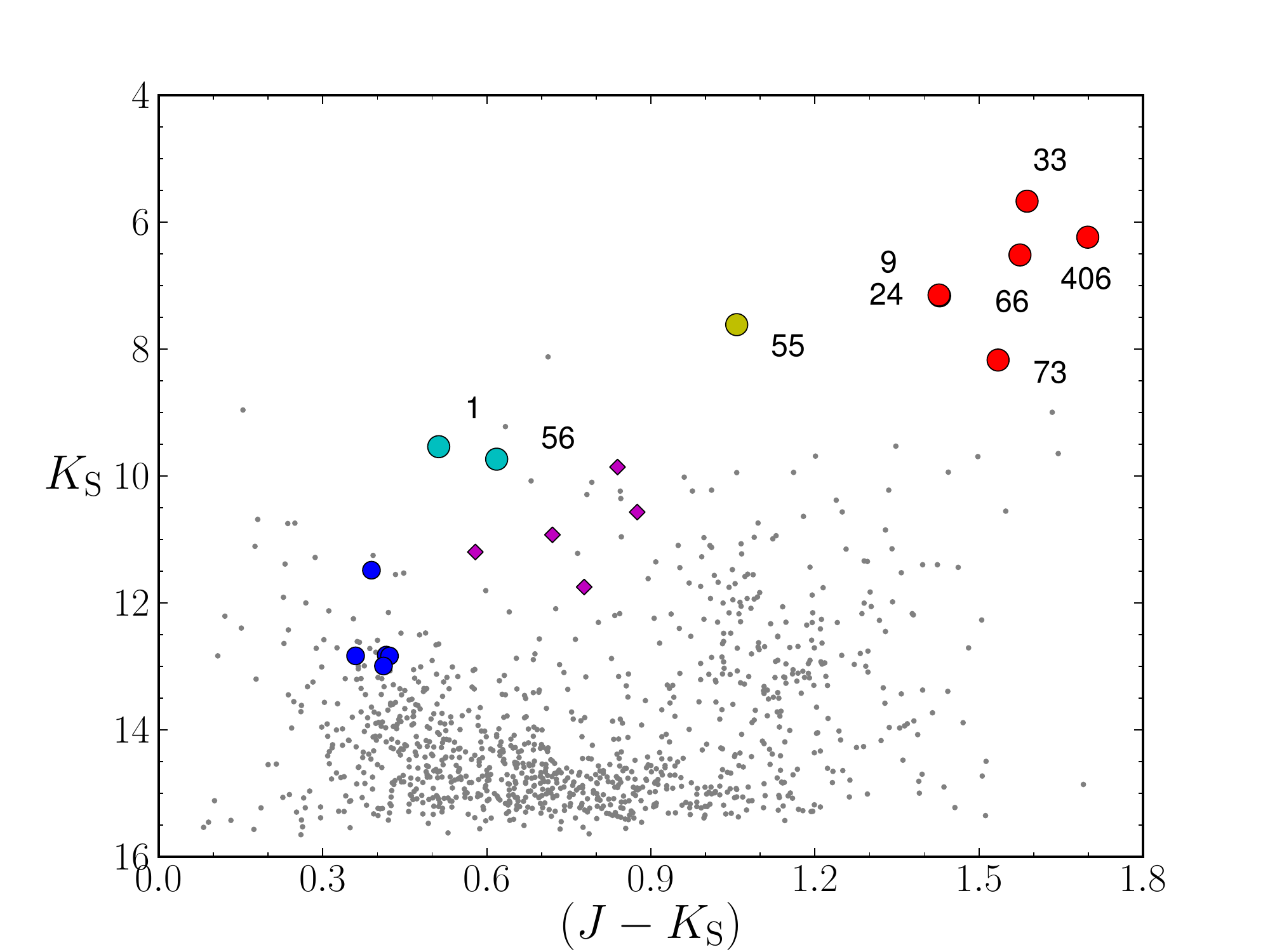}   
  \caption{$K_{\textrm{S}}$/$(J-K_{\textrm{S}}$) diagram for all stars with good 2MASS data in the field of NGC~3105. Symbols and colours follow the same code as in Fig.~\ref{CMD_opt}.
  Stars 9 and 24 occupy the same position on the diagram. We note the location of S406, a likely new cluster RSG, which is not covered by the optical photometry.} 
  \label{CMD_jhk}  
\end{figure}

In order to complement our photometry we resorted to the 2MASS photometry. In Fig.~\ref{CMD_jhk}, we also show the $K_{\textrm{S}}$/$(J-K_{\textrm{S}})$ CMD from the 2MASS data for stars located up to $5\arcmin$
from the nominal centre. Because of its high distance modulus the top of the MS is found only one magnitude and a half above the completeness
limit in the $K_{\textrm{S}}$ band ($\sim$14.3). Thus, the cluster is too faint for performing a complete analysis with the 2MASS data, analogous to 
that carried out with the optical photometry. However, this diagram (Fig.~\ref{CMD_jhk}) shows the position of star 406, a possible new RSG (whose cluster memberhip has not been claimed so far), not covered by the optical photometry.

\subsubsection{Reddening}
Firstly, we selected the B-type stars spectroscopically identified in that sequence (Fig.~\ref{CMD_opt}) without emission lines or red companions. For these stars, six in total, we determined their individual reddening by comparing 
the observed $(B-V)$ colours with the $(B-V)_0$ corresponding to their spectral types (Table~\ref{reddening}). We make a first estimate of the cluster reddening by taking an average of the values for these stars. 
We obtain a colour excess $E(B-V)\,=\,1.03\,\pm\,0.03$. The low dispersion suggests an absence of strong differential reddening that allows a meaningful determination of the average. In order to estimate whether the extinction law is standard in this direction, we also calculate for these six stars
the ratio $X\,=\,E(U-B)\,/\,E(B-V)$. We find an average $X=0.76\pm\,0.11$, compatible with the standard value of 0.72 \citep{JM53}. In both cases the intrinsic colours are adopted from \citet{Fi70} and the errors show the
dispersion between different stars. 

\begin{table*}
  \caption{Colour excesses and absolute magnitudes in the $V$ band for B-type stars in the field of NGC~3105. Intrinsic colours are adopted from \citet{Fi70}.\label{reddening}}
\begin{center}
\begin{tabular}{cccccccccc}
\hline\hline
NGC~3105 & Sp T & $(B-V)$ & $(U-B)$ & $(B-V)_0$ & $(U-B)_0$ & $E(B-V)$ & $E(U-B)$ & $X$ & $M_V$\\
\hline
12 & B2\,V   & 0.790 & $-$0.017 & $-$0.24 & $-$0.81 & 1.030 & 0.793 & 0.770 & $-$2.31 \\
14 & B3\,III & 0.785 & $-$0.004 & $-$0.20 & $-$0.73 & 0.985 & 0.726 & 0.737 & $-$3.55 \\
16 & B2.5\,V & 0.808 &    0.060 & $-$0.22 & $-$0.72 & 1.028 & 0.780 & 0.759 & $-$2.19 \\
25 & B2\,V   & 0.786 & $-$0.020 & $-$0.24 & $-$0.81 & 1.026 & 0.790 & 0.770 & $-$2.20 \\
34 & B2.5\,V & 0.798 & $-$0.003 & $-$0.22 & $-$0.72 & 1.018 & 0.717 & 0.704 & $-$2.12 \\
71 & B2\,IV  & 0.828 & $-$0.008 & $-$0.24 & $-$0.86 & 1.068 & 0.852 & 0.798 & $-$1.60 \\

\hline 
\end{tabular}
\end{center}
\end{table*}

\subsubsection{Cluster membership and fitting the ZAMS}\label{sec_miem}

Since blue members in stellar clusters are easy to distinguish from the field population, and hence good tracers of the cluster extent, we looked for them covering the range from the
earliest objects found with spectral types B2-2.5 down to the spectral type B9, for which we have completeness in our photometry as mentioned before (Sect.~\ref{general}).
Since the extinction is close to standard, we can use the classical $Q$ parameter \citep{JM53}, defined as $Q=(U-B)-X\,(B-V)$. This parameter, which is corrected from interstellar reddening, allows efficient separation of blue stars from the rest
as well as assigning reliable photometric spectral types. 
From photometry and the average $X$ ratio found in the 
previous subsection we calculated the value of $Q$ for the totality of objects that we have. In order to find stars intrinsically blue along the cluster field, we chose stars with $-0.74<Q<-0.06$ as described in \citet{JM53}.
Then, we assigned photometric spectral types to these stars and selected as likely members those whose photometric spectral types are in fair agreement with their expected location on the CMD (Fig.~\ref{CMD_opt}). 
For the early-type stars with spectra, the difference between spectroscopic and photometric spectral types is typically only one subtype.
Finally we find 126 likely members, whose individual photometric reddening can be estimated using this expression \citep{john58}:

\begin{equation}
E(B-V)\,=\,(B-V)\,-\,0.332\,Q 
\end{equation}

For these stars, the average reddening is $E(B-V)\,=\,1.01\,\pm\,0.04$, fully compatible with that obtained for the limited sample of objects with spectra using the colour calibration. In Table~\ref{phot_spt}, we list the likely
members together with their $Q$ index, photometric spectral type and reddening estimated in this section. In addition, the last column of Table~\ref{SpT} summarises the cluster membership
for all stars observed spectroscopically. Most of them are likely blue members as just discussed. The criteria for selecting red members are based on the position on CMD (Sect.~\ref{iso_3105}) as well as the analysis of radial velocities (Sect.~\ref{sec:rv}).

\begin{figure}  
  \centering         
  \includegraphics[width=\columnwidth]{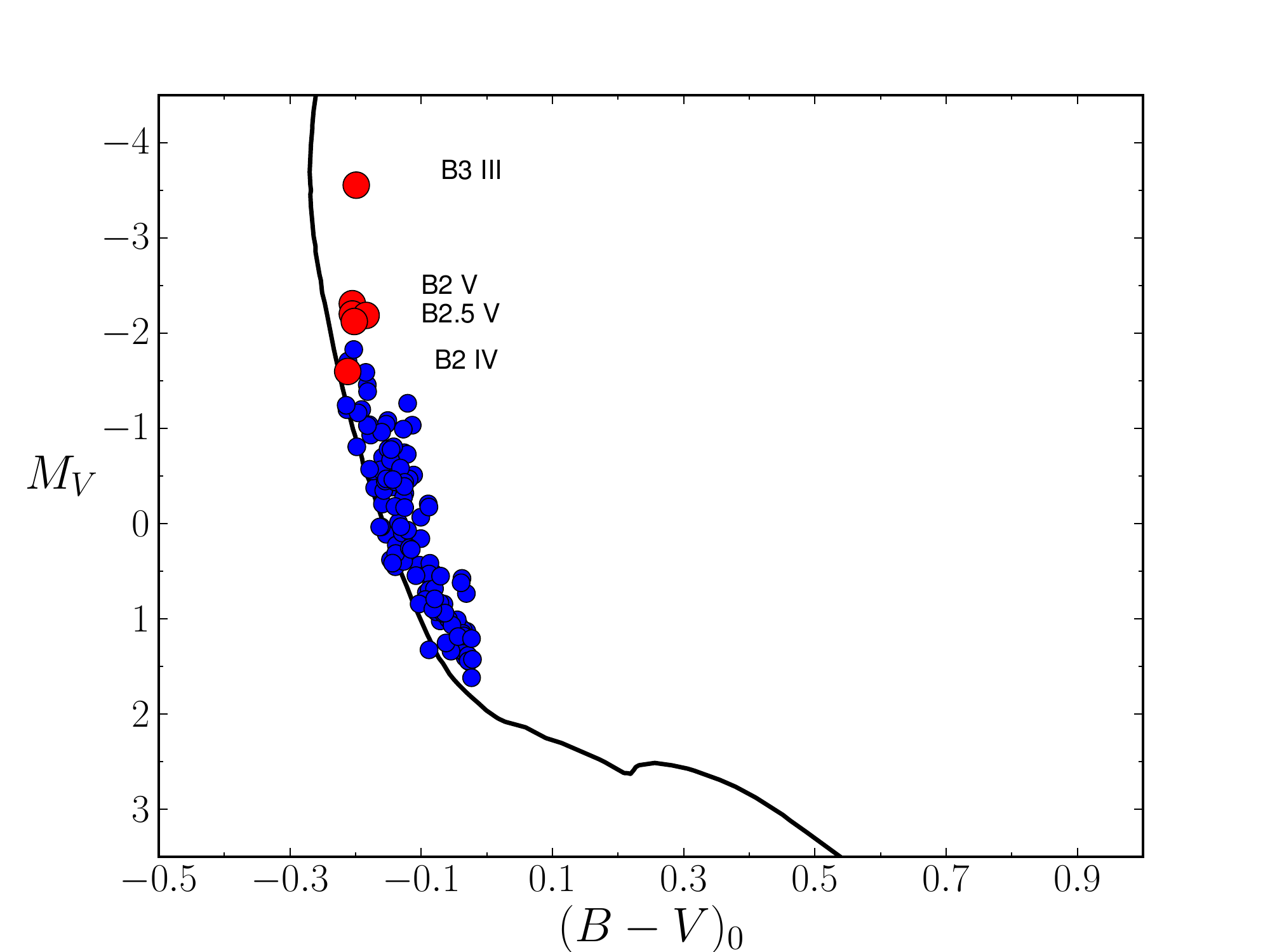}   
  \caption{$M_V/(B-V)_0$ diagram for likely members of NGC~3105. Stars observed spectroscopically are represented in red colour. Spectral types for the brightest stars are marked. Blue circles are the rest of early members found photometrically. 
  The black line shows, as a ZAMS, a PARSEC isochrone of 10\,Ma computed at the cluster metallicity. The best-fitting yields a distance modulus, $\mu=14.3\pm0.2$.} 
  \label{zams}  
\end{figure}

Once we know the individual reddening for likely blue members, we carefully performed a visual fit of these dereddened values with the observational zero age main sequence (ZAMS).
Classical ZAMSs \citep[such as that from][]{SK82} have been calibrated with bright stars in the solar neighbourhood, easily observable, with typically solar abundances. Since this cluster exhibits a lower metallicity, it is necessary to employ a more appropriate ZAMS. To this end, we computed
a PARSEC isochrone with the cluster metallicity, but with an age young enough (i.e. 10\,Ma) compared to that of the cluster itself.
From the ZAMS fitting (see Fig.~\ref{zams}) we obtain the distance modulus, $\mu=V_0-M_V=14.3\pm0.2$, which is equivalent to a distance $d=7.2\pm0.7\:$kpc. The error reflects the uncertainty when visually fitting the ZAMS as a lower envelope.
This distance places the cluster with respect to the Galactic centre, taking as solar reference $R_{\sun}=8.3\:$kpc, at $R_{\textrm{GC}}=10.0\pm0.9\:$kpc.

Finally, we take advantage of the combination of photometry with spectroscopy to check the validity of our result. In this way, the $M_V$ resulting for stars with spectra are compatible with those 
expected according to their spectral types.

\subsubsection{Fitting isochrones}\label{iso_3105}
We employed the isochrone fitting method to determine the age of the cluster. Initially, taking into account the distance modulus just estimated, we plotted the $M_V/(B-V)_0$ diagram and then, on it, we fitted by eye PARSEC isochrones \citep{Parsec} 
computed at the cluster metallicity. The metallicity, obtained in this work (See Sect.~\ref{sec_par_cool}), in terms of iron abundance is [Fe/H]\,=\,$-$0.29. We converted this value into $Z$ by using 
the approximation [M/H]\,=log($Z/Z_{\sun}$), with $Z_{\sun}$=0.0152 for PARSEC tracks. 

\begin{figure}  
  \centering         
  \includegraphics[width=\columnwidth]{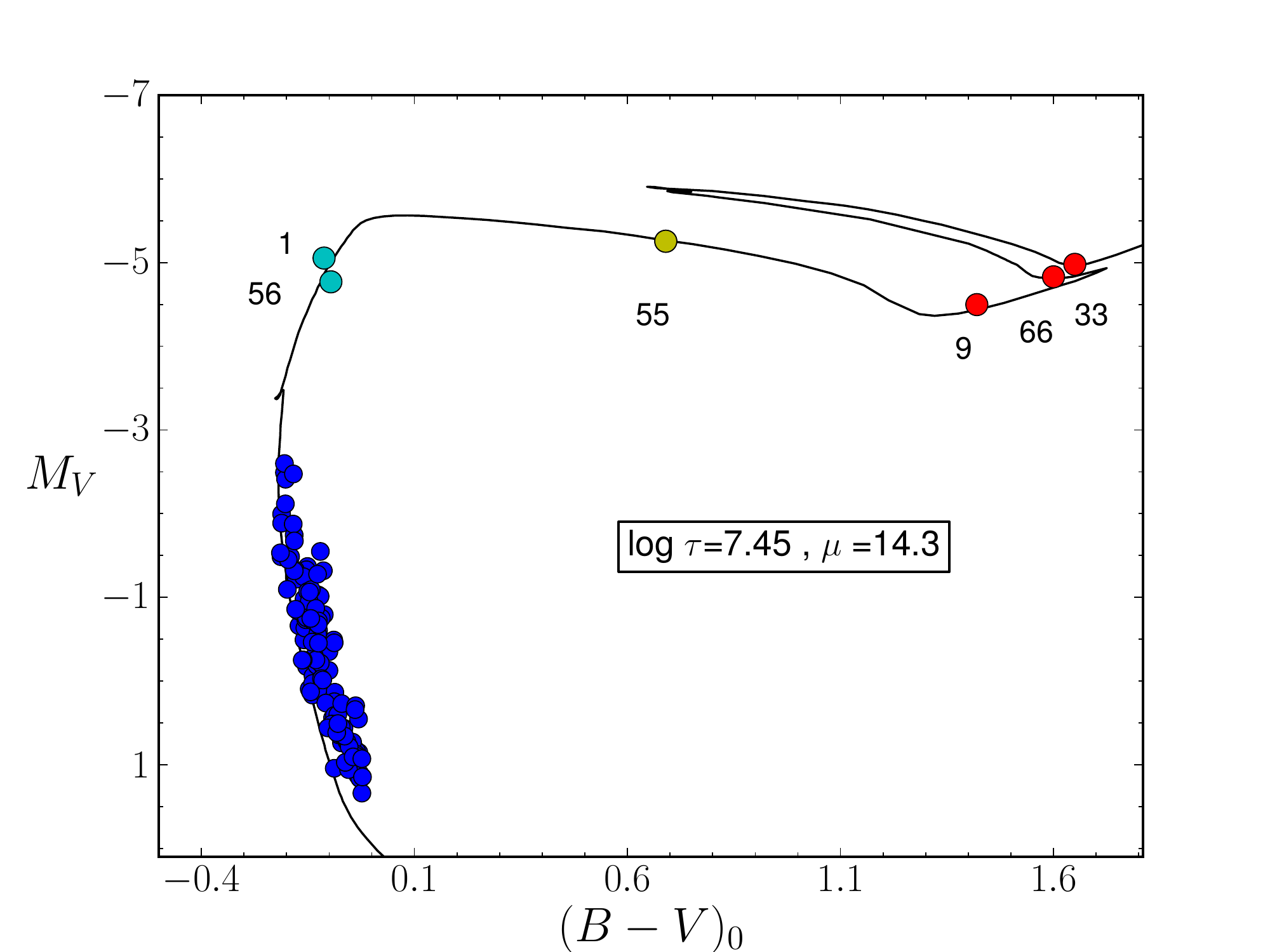}   
  \caption{$M_V/(B-V)_0$ diagram for likely members of NGC~3105. Symbols and colours follow the same code as in Fig.~\ref{CMD_opt}. The best-fitting isochrone is the black line.}
  \label{iso_opt_3105}  
\end{figure}

Figure~\ref{iso_opt_3105} shows the optical dereddened CMD, $M_V/(B-V)_0$, for likely members and the best-fitting isochrones. The dereddening of the blue members has been described before.
For all supergiants we followed the procedure described by \citet{fernie63}. Only single stars without emission lines are included.
Although we do not provide photometry for S33 we use the value given by \citet{Fi77} taking into account that there is no offset 
between both photometries (Table~\ref{dif_fotom}). When fitting the isochrones, it is key to pay special attention to the position 
of evolved members, from which we infer the cluster age. 
The best-fitting PARSEC isochrone gives a log\,$\tau$\,=\,7.45$\pm$0.10, where the error represents the typical range of isochrones which give a good fit. 
At this age, the mass of the RSGs would correspond to 9\,--\,10\:M$_{\sun}$, somewhat smaller than that expected from previous works.

\subsection{Cluster geometry}\label{sec_geom_3105}
\begin{figure}  
  \centering         
  \includegraphics[width=\columnwidth]{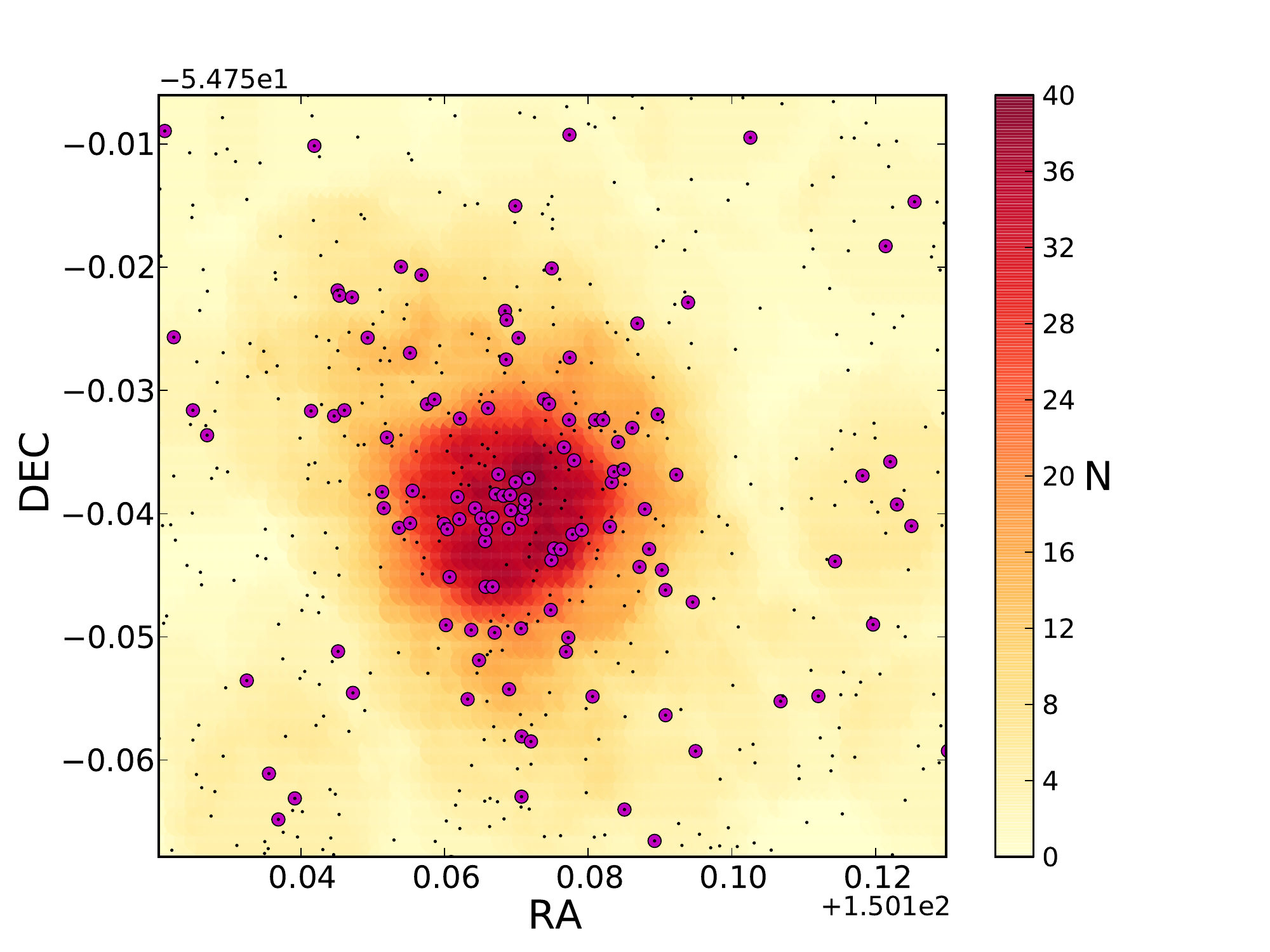}   
  \caption{Sky distribution of the likely members of NGC~3105 (magenta circles) and the rest of observed stars (black dots) in the field covered by our photometry. Density is computed as the number of 
  stars closer than $0\farcm5$ to each point. The overdensity corresponding to the cluster core clearly stands out in the figure.}
  \label{gausiana_3105}  
\end{figure}

From the sky density distribution of the 126 likely members the centre and size of the cluster are derived as described in \citet{7067}. For this purpose the observed sky field has  been divided symmetrically
into a grid of points spaced $5\arcsec$. The sky density has been computed as the number of stars located closer than $0\farcm5$ from each point of this grid. The cluster is enhanced from the 
background and the overdensity at its central part is clearly detected (see Fig.~\ref{gausiana_3105}). Then, we fit this density with a 2-D Gaussian such as:

\begin{equation}
n = n_0 + n_{\textrm{max}}\,\cdot\,e^{-(A(\alpha-\alpha_0)^2+2B(\alpha-\alpha_0)(\delta-\delta_0)+C(\delta-\delta_0)^2)}
\end{equation}
with
\begin{gather*}
A = \frac{\cos^2\theta}{2\sigma^2_{\alpha}} + \frac{sin^2\theta}{2\sigma^2_{\delta}}  \\   
B = \frac{-sin(2\theta)}{4\sigma^2_{\alpha}} + \frac{sin(2\theta)}{4\sigma^2_{\delta}}  \\ 
C = \frac{sin^2\theta}{2\sigma^2_{\alpha}} + \frac{cos^2\theta}{2\sigma^2_{\delta}}
\end{gather*}

where there are seven parameters to fit: $n_0$ is the offset due to the zero point, $n_{\textrm{max}}$ is the amplitude of the Gaussian, $\alpha_0$ and $\delta_0$ are the equatorial coordinates
of the cluster centre, $\sigma_{\alpha}$ and $\sigma_{\delta}$ are the standard deviation corresponding to each coordinate and $\theta$ is the position angle of the ellipse with respect to the vertical.
The results, obtained through a least-squares fit, are listed in Table~\ref{geom_3105}.

\begin{table}
\caption{Parameters obtained for the cluster geometry.}
\begin{center}
\begin{tabular}{lc}   
\hline\hline
Parameter          & Value\\
\hline
$n_0$              & $2.286\,\pm\,0.009$ stars/arcmin$^2$\\
$n_{\textrm{max}}$ & $35.80\,\pm\,0.19$ stars/arcmin$^2$\\
$\alpha_0$         & $150.17020\,\pm\,0.00007\,\degr$\\
$\delta_0$         & $-54.78922\,\pm\,0.00005\,\degr$\\
$\sigma_{\alpha}$  & $0.01397\,\pm\,0.00008\,\degr$\\
$\sigma_{\delta}$  & $0.00915\,\pm\,0.00005\,\degr$\\
$\theta$           & $4.8\,\pm\,0.5\,\degr$\\
\hline
\end{tabular}

\label{geom_3105}
\end{center}
\end{table}

The cluster extends across the sky projecting an ellipse whose eccentricity is given by the expression:
\begin{equation}
 \epsilon\,=\,\sqrt{1-\dfrac{(\sigma_{\alpha}\,cos\,\delta_0)^2}{\sigma^{2}_{\delta}}}\,=\,0.51    
\end{equation}

Finally, we can define the radius of the cluster ($r_{\textrm{cl}}$) as the distance from its centre at which almost all the cluster members are contained (within $3\sigma$):

\begin{equation}
r_{\textrm{cl}}\,=\,3\,\sqrt{(\sigma_{\alpha}\,cos\,\delta_0)^2+\sigma_{\delta}^2} 
\end{equation}

According to this definition, and taking into account the uncertainties in each of the parameters, we estimate a cluster radius, $r_{\textrm{cl}}=2\farcm7\pm0\farcm6$, that at the distance
of the cluster corresponds with a physical size of $5.8\pm1.7$~pc. This result is consistent with the position of stars 66 and 406. Among those members
confirmed via radial velocity, both stars are the most distant objects from the cluster centre, located around $2\arcmin$, within the radius found in this work.
Star 101 lies in the cluster surroundings, at $5\arcmin$, beyond the cluster radius. This implies that this star either is not a cluster member or is placed in the cluster outer halo.

\subsection{Cluster mass}
Once the size of the cluster is fixed, we continue our analysis by estimating its mass using the initial mass function (IMF). For this purpose we employed the multiple-part power law IMF defined 
by \citet{Kroupa}.

\begin{equation}
\xi(M) = \xi_0\,\cdot\,M^{-\alpha_i} \left\lbrace
\begin{array}{lll}
\alpha_1 = 1.3, \hspace{5mm} 0.08 \leq M/\textrm{M}_{\sun} < 0.50\\
\alpha_2 = 2.3, \hspace{5mm} 0.50 \leq M/\textrm{M}_{\sun} 
\end{array}
\right.
\end{equation} 

where $\xi_0$ is the constant which sets the local stellar density. 

By integrating the IMF over a mass range between $M_1$ and $M_2$, the number of stars ($N$) expected in this interval is given by:

\begin{multline}
N = \int_{M_1}^{M_2} \xi(M)\,\textrm{d}M = \xi_0\,\int_{M_1}^{M_2} M^{-\alpha_i}\,\textrm{d}M \\
\Rightarrow \hspace{3mm}\xi_0 = \frac{N}{\int_{M_1}^{M_2} M^{-\alpha_i}\,\textrm{d}M}
\end{multline}

Firstly, we set the free parameter of the IMF ($\xi_0$) by counting the stars within a certain mass range. We chose B stars ranging from spectral 
type B2\,V (which corresponds to the spectral type of the earliest members) to B9\,V, where the separation from the field population is very efficient and completeness is ensured (Sect.~\ref{general}).
According to the calibration of \citet{st92}, this range of spectral types covers stars between 2.6 and $9.8\:$M$_{\sun}$. 
These stars, all inside the cluster radius, are the 126 likely members, previously selected in Sect.~\ref{sec_miem}. 

Once the value of $\xi_0$ is known ($\xi_0\approx707$), the total mass of the cluster can be estimated using the following expression:

\begin{equation}
M = \int_{M_1}^{M_2} M\,\xi(M)\,\textrm{d}M = \xi_0\,\int_{M_1}^{M_2} M^{1-\alpha_i}\,\textrm{d}M
\end{equation}

Then, integrating between 0.08 and 9.5\,M$_{\sun}$ (the mass of the evolved stars according to the isochrone) we obtain the present mass, whereas integrating up to 150\,$M_{\sun}$ (as a classical estimate of the maximum mass for a star
\footnote{\citet{r136a1} raise this value above 300\,M$_{\sun}$. In any case this number, if confirmed as the actual initial mass of a single star or as the resulting mass of a stellar merger,
hardly affects the cluster mass.}) we determine the
initial mass. For NGC~3105 we find $\approx$2\,300 and $\approx$3\,000\,M$_{\sun}$ respectively. 
Finally, according to \citet{Tadross} we correct these values for unresolved binarity by assuming an average mass ratio of binary systems of 1.3 \citep{Allen} and a binary frequency of 50\,$\%$ \citep{Ja70}. 
\citet{Kouwenhoven05} found a fraction around 80\,$\%$ (B0-B3) and 50\,$\%$ (B4-B9) among B stars in the Sco OB2 association. Recently, \citet{Dunstall15}, in the 30 Dor star forming region, also estimated an observed mean binary fraction
around 25\,$\%$ but the implied underlying (true) value (according to their simulations) is 58\,$\%$, a fraction very similar to our assumption (50$\%$).
We obtain a total present mass of $\approx$3\,200\,M$_{\sun}$, and a total initial mass around 4\,100\,M$_{\sun}$.
Hence, we found that NGC~3105 is a moderately massive cluster.

\subsection{Rotational and radial velocities}\label{sec:rv}

As mentioned before (Sect.~\ref{espectros}) we took low- and moderate-resolution spectra for blue stars. This resolution is not as high as that necessary to separate different broadenings and 
we assumed that the whole broadening has a rotational origin. Starting from a first estimation of $v\sin\,i$ (around 50\,km\,s$^{-1}$, a value close to the resolution) an initial model capable of reproducing the 
spectrum was computed. In an iterative way, once we set the stellar parameters, we looked for a second estimate of $v\sin\,i$ by changing it until finding a new model which reproduced best the profiles. This process
was repeated a couple of times until it did not change the rotational velocity. The values obtained for $v\sin\,i$, together with the atmospheric parameters, are listed in Table~\ref{Par_cal_3105}.

In contrast, for the six stars observed at high resolution with FEROS we computed $v \sin\,i$ by using the {\scshape iacob-broad} code \citep{iacob_broad}, based on the Fourier transform method, which separates rotation
from other broadening mechanisms such as the macroturbulent velocity ($\zeta$). We used six lines of \ion{Fe}{i} and \ion{Ni}{i}. The errors reflect the scatter between measurements, in terms of rms. 
Results are shown in Table~\ref{Par}.

For these stars we also obtained RVs, referred to the heliocentric reference frame of rest, through Fourier cross-correlation by employing the software {\scshape iSpec} \citep{ispec}, especially designed for the study of 
cool stars. The cross-correlation is computed against a list of atomic line masks, carefully selected for the {\em Gaia} benchmark stars library pipeline, from asteroids observed with the NARVAL spectrograph.

Star 73 (See Table~\ref{Par}) presents a discrepant RV with respect to the rest, suggesting that it is a field star. This is confirmed by its low luminosity class. The remaining objects have very similar radial velocities; star 24 shows a RV slightly lower than the other 
supergiants, but such a discrepancy is fully consistent with its binary nature. Also star 406 shows a RV compatible with other supergiants, but because of its uncertain nature (see discussion in Sect.~\ref{406}) it has not been used to calculate the cluster average.
Thus, we can use the other three stars (namely 9, 55 and 66) to estimate the RV for the cluster by averaging the individual values, obtaining a $v_{\textrm{rad}}$=$+46.9\pm0.9\:\textrm{km}\,\textrm{s}^{-1}$,
where the error is the dispersion between the values for the different stars.

\subsection{Stellar atmospheric parameters}\label{sec_param}

\subsubsection{Blue stars}
From low-medium resolution EFOSC2 spectra ($R\sim$1\,000) we computed stellar parameters for 12 blue stars. When possible we used the spectra taken with grism 19, since it provides a higher resolution than grism 14.

We employed the technique described by \citet{Ca12}, also in \citet{Le07}. The stellar atmospheric parameters were derived through an automatic $\chi^2$-based algorithm searching for
the set of parameters that best reproduce the main strong lines observed in the range $\approx$\,4000\,--\,5000\,\AA{}. 
We used a grid of {\scshape fastwind} synthetic spectra \citep{Si11,Ca12}. The stellar atmosphere code {\scshape fastwind} \citep{Sa97,Pu05} enables non-LTE calculations and assumes a spherical geometry as well as an explicit treatment 
of the stellar wind effects. 
The results are shown in Table~\ref{Par_cal_3105}. 
It should be noted that the emission present in the Be stars has prevented a good fit with the models, yielding unrealible parameters, such as gravities smaller than
those computed for supergiants. For this reason, we ignore the results for these objects. Temperatures for non-emission B stars are fully compatible with those from the 
calibration by \citet{Hump84}. The temperatures for A0 supergiants are somewhat hotter than the calibration.

\begin{table*}
\caption{Log of EFOSC2 spectra and stellar atmospheric parameters for blue stars with no emission lines and stellar atmospheric parameters derived from them. \label{Par_cal_3105}}
\begin{center}
\begin{tabular}{lccccccc}   
\hline\hline
Star   & Sp T & Grism & $t_{\textrm{exp}}$ (s) & $S/N$ & $v \sin\,i$ (km\,s$^{-1}$) & $T_{\textrm{eff}}$ (K) & $\log\,g$ \\
\hline
1   & A0\,Ib  & 19 & 900    &  130  &  90  & 12\,000 $\pm$ 1\,000 & 2.6 $\pm$ 0.1 \\
12  & B2\,V   & 14 & 1\,800 &  60   & 110  & 18\,000 $\pm$ 2\,200 & 3.8 $\pm$ 0.2 \\
14  & B3\,III & 19 & 1\,800 &  90   &  90  & 17\,000 $\pm$ 1\,300 & 3.1 $\pm$ 0.2 \\
16  & B2.5\,V & 14 & 1\,800 &  160  & 230  & 17\,000 $\pm$ 1\,000 & 4.1 $\pm$ 0.1 \\
25  & B2\,V   & 14 & 1\,800 &  60   & 230  & 18\,000 $\pm$ 1\,000 & 4.1 $\pm$ 0.1 \\
34  & B2.5\,V & 14 & 1\,800 &  60   & 210  & 16\,000 $\pm$ 1\,000 & 3.8 $\pm$ 0.1 \\
56  & A0\,Ib  & 19 & 1\,200 &  120  &  70  & 12\,000 $\pm$ 1\,000 & 2.6 $\pm$ 0.1 \\
71  & B2\,IV  & 14 & 1\,800 &  50   & 230  & 20\,000 $\pm$ 1\,600 & 4.1 $\pm$ 0.2 \\
\hline

\end{tabular}
\end{center}

\end{table*}

\subsubsection{Cool stars}\label{sec_par_cool}

For the cool stars we derived stellar atmospheric parameters from their FEROS spectra. 
For the three K-supergiants and the yellow supergiant we followed the procedure described in \citet{6067}, but using different atmospheric models. In this work we took 1D LTE atmospheric models, specifically
MARCS spherical models with $1\:$M$_{\sun}$ \citep{marcs}. We generated a grid of synthetic spectra by using the radiative transfer code {\scshape spectrum} \citep{graco94}. Although MARCS models 
are spherical, {\scshape spectrum} treats them as if they were plane-parallel. Therefore, the plane parallel transfer treatment might produce a small inconsistency in the calculations of synthetic 
spectra based on MARCS atmospheric models. However, the study of \citet{hei06} concluded that any difference introduced by the spherical models in a plane-parallel transport scheme is small. 
The microturbulent velocity ($\xi$) was fixed according to the calibration given in \citet{dutrafe16}. Effective temperature (T$_{\textrm{eff}}$) ranges from 3\,300~K to 7\,500~K with a step of 100~K up to 4\,000~K and 
250~K until 7\,500~K, whereas surface gravity ($\log\,g$) varies from $-0.5$ to 3.5~dex in 0.5~dex steps. Finally, in the case of metallicity (using [Fe/H] as a proxy), the grid covers from $-1.5$ to
$+1.0$~dex in~0.25 dex steps.

We employed a methodology to derive stellar atmospheric parameters based on the iron linelist provided by \citet{Ge13}, since Fe lines are numerous as well as very sensitive in cool stars. 
The linelist contains $\sim$230 features for \ion{Fe}{i} and $\sim$55 for \ion{Fe}{ii}. Their atomic parameters were taken from the VALD database\footnote{\url{http://vald.astro.uu.se/}} \citep{pis95,kup2000}. For the Van der Waals 
damping data, we took the values given by the Anstee, Barklem, and O'Mara theory, when available in VALD \citep[see ][]{bar00}. 

\begin{figure}  
  \centering         
  \includegraphics[width=\columnwidth]{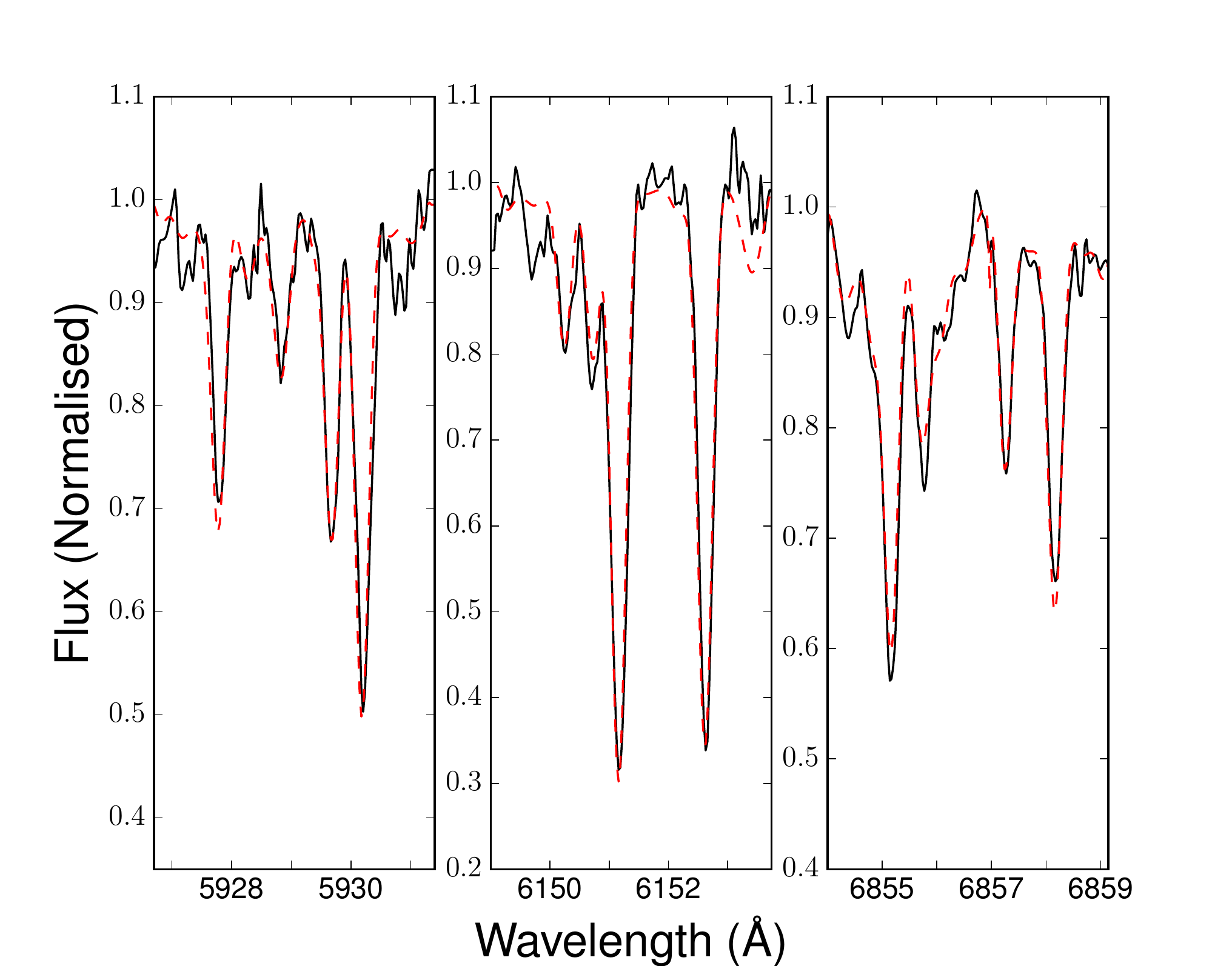}   
  \caption{Some examples of line fits in star 66. Three regions are shown around \ion{Fe}{i}\,$\lambda$5927.80 (left), \ion{Fe}{ii}\,$\lambda$6149.25 (centre) and 
  \ion{Fe}{i}\,$\lambda$6855.74--\ion{Fe}{i}\,$\lambda$6858.16 (right).
 The observed spectrum is the solid black line, whereas the synthetic one is the red dashed line.} 
  \label{ajuste_lineas}  
\end{figure}

As a starting point, we employed an updated version of the {\scshape stepar} code \citep{hugo18},   
adapted to the present problem, that uses stellar synthesis instead of an $EW$ method. 
As optimization method we used the Metropolis--Hastings algorithm.
Our method generates simultaneously 48 Markov-chains of 750 points each one. It then performs a Bayesian parameter estimation by employing an implementation of Goodman $\&$ Weare's Affine Invariant Markov chain Monte Carlo Ensemble sampler
\citep{mcmc}. 
As objective function we used a $\chi^2$ in order to fit the selected iron lines. We fixed the stellar rotation to that value previously derived, and the instrumental broadening to the resolution of the FEROS spectrograph. 
We left the macroturbulent broadening as a free parameter to absorb any residual broadening. An example of the spectral line fitting performed with our methodology is shown in Fig.~\ref{ajuste_lineas}.

In the case of the M-(super)giants, since their spectra are dominated by molecular bands which erode the continuum weakening or erasing other spectral features such as most iron lines, we were
forced to modify our methodology. We followed the procedure described in \citet{anib07}, focusing in the region 6670--6730\,\AA{} where the TiO bands at 6681 and 6714\,\AA{} are clearly present. These bands are very useful since 
their depth is very sensitive to temperature \citep{anib07}. We used the same grid of MARCS synthentic spectra but using {\scshape turbospectrum} \citep{turbo} as a transfer code. Due to the strong degeneracy between metallicity 
and gravity we fixed the former with the mean value derived from the FK supergiants. This is valid when the M star has a RV compatible with the cluster average. 
If not, we assumed a log\,$g$=0.0, leaving the $T_{\textrm{eff}}$ as a free parameter.

Results (i.e. effective temperature, surface gravity, macroturbulent velocity and iron abundance) are displayed in Table~\ref{Par}.
From the analysis of the three K and the F supergiant, we derive a subsolar metallicity for the cluster. The weighted arithmetic mean (using the variances as weights) is [Fe/H]$=-0.29\pm0.22$.
The uncertainty, quite conservative, is expressed in terms of the standard deviation around this mean. We find a relatively high dispersion between our values for metallicity. Given that the star with higher metallicity, S24,
has a blue companion that may affect the calculation, the actual cluster metallicity could be even lower.

\begin{table*}
\caption{Log of FEROS spectra and stellar atmospheric parameters for the cool stars derived from them. \label{Par}}
\begin{center}
\begin{tabular}{lccccccccc}   
\hline\hline
Star & Sp T & $t_{\textrm{exp}}$ (s) & $S/N$ & $v_{\textrm{rad}}$ (km\,s$^{-1}$) & $v \sin\,i$ (km\,s$^{-1}$) &  $\zeta$ (km\,s$^{-1}$) & $T_{\textrm{eff}}$ (K) & $\log\,g$ & [Fe/H]\\
\hline
\multicolumn{10}{c}{Members}\\
\hline
9   & K3\,Ib & 3\,600 & 80  & 47.16 $\pm$ 0.03 & 6.6 $\pm$ 1.4 & 1.56 $\pm$ 0.90  & 4\,038 $\pm$ 134 & 0.78 $\pm$ 0.39 & -0.47 $\pm$ 0.18\\
24$^{*}$ & K3\,Ib & 3\,600 & 80  & 40.60 $\pm$ 0.04 & 7.4 $\pm$ 1.1 & 7.50 $\pm$ 0.25  & 4\,285 $\pm$ 80  & 1.42 $\pm$ 0.22 & -0.20 $\pm$ 0.11\\
55  & F9\,Ib & 3\,600 & 70  & 45.48 $\pm$ 0.07 & 7.0 $\pm$ 1.6 & 12.61 $\pm$ 1.11 & 5\,926 $\pm$ 141& 1.57 $\pm$ 0.30 & -0.24 $\pm$ 0.10\\ 
66  & K5\,Ib & 3\,600 & 100 & 46.82 $\pm$ 0.03 & 7.8 $\pm$ 1.4 & 3.66 $\pm$ 1.05  & 3\,930 $\pm$ 147 & 0.29 $\pm$ 0.43 & -0.55 $\pm$ 0.19\\
406 & M6\,Ib & 2\,700 & 40  & 47.15 $\pm$ 0.07 & 9.7           &  1.90            & 3\,300 $\pm$ 100 & 0.00 $\pm$ 0.50 & $-$0.29 \\
\hline
\hline
\multicolumn{10}{c}{Non-member}\\
\hline
73  & M2\,III& 5\,400 & 60 & 59.38 $\pm$ 0.04 & 7.5  &  7.1  & 3\,600 $\pm$ 100 & 0.00 & $-$0.10 \\
\hline

\end{tabular}
\end{center}
\begin{list}{}{}
\item[]$^{*}$ This star exhibits a composite spectrum, whose companion is a blue star (B2\,V:).
  \end{list}
\end{table*}

\begin{figure}  
  \centering         
  \includegraphics[width=\columnwidth]{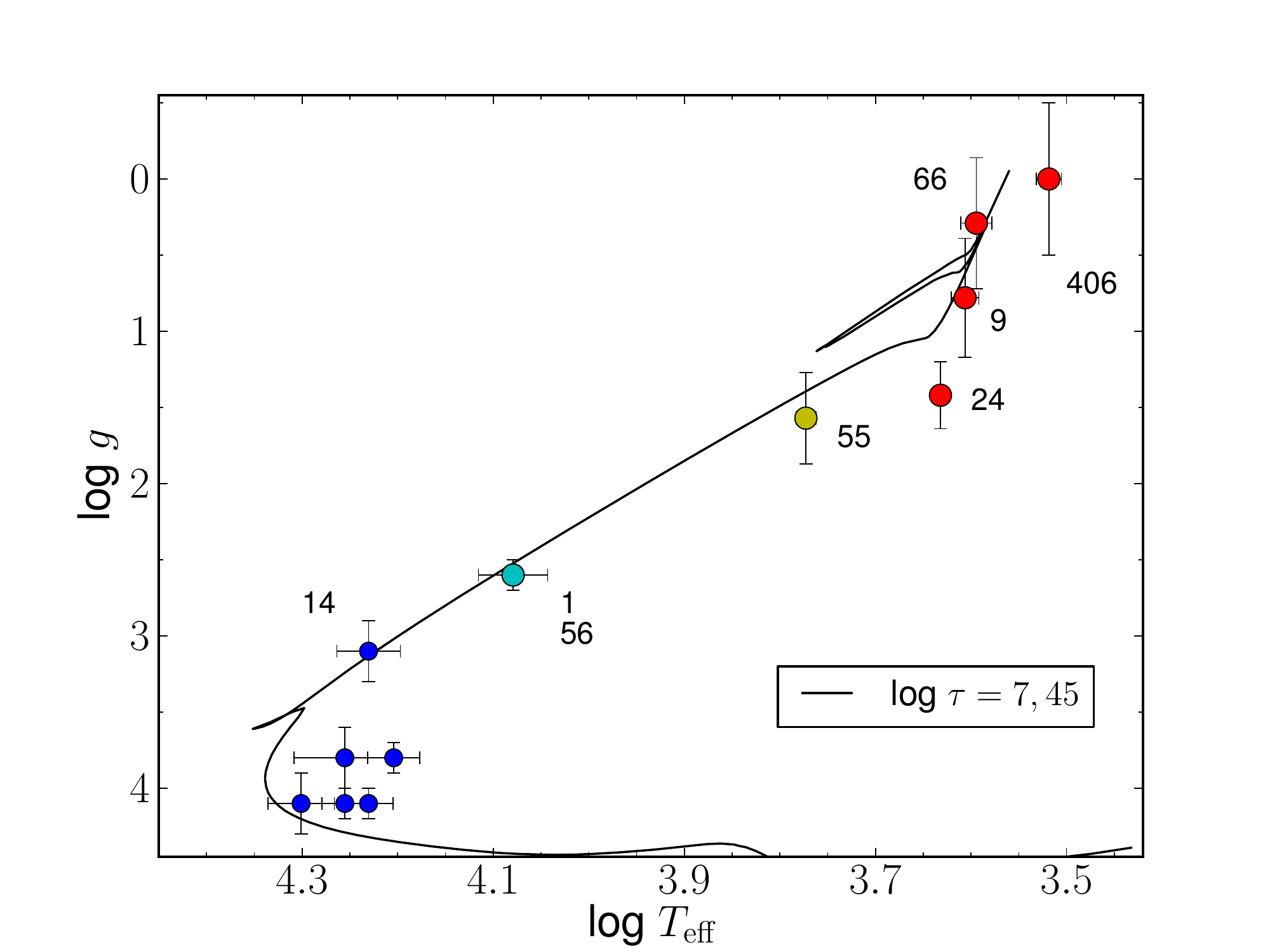}   
  \caption{Kiel diagram for likely members (with the exception of Be stars). Symbols and colours follow the same code as in Fig.~\ref{CMD_opt}. 
  Stars 1 and 56 occupy the same position on the diagram.} 
  \label{pHR_3105}  
\end{figure}

Finally, in Fig.~\ref{pHR_3105} we plot the Kiel diagram (i.e. $\log\,{g}$\,--\,$\log\,T_{\textrm{eff}}$) for likely members for which we have stellar parameters (Tables~\ref{Par_cal_3105} and \ref{Par}).
On this diagram, independent of the cluster distance, we also plot the best-fitting isochrones according to that seen in Fig.~\ref{iso_opt_3105}. The isochrone, with the expected exceptions of the Be stars (7 and 72),
reproduces quite good the observed evolutionary location of stars. It should be noted that supergiant stars (blue, yellow and red) lie all very close to the isochrone.
Only S24 (composite spectrum) and S406 (see further discussion in Sect.~\ref{406}) fall away from it.

The good accordance between spectral results (stellar parameters) and cluster parameters such as the distance and age inferred photometrically supports the reliability of our results. 

\subsection{Chemical abundances}

We derived chemical abundances for the three K-supergiants, since for the remaining stars (including the YSG)
there are no spectral diagnostic features with enough quality for their analysis. 

We employed the same methodology as in the study of NGC 6067 \citep{6067}. We used a method based on $EW$s measured in a semi-automatic fashion using {\scshape tame} \citep{kan12}
for Na, Mg, Si, Ca, Ti, Ni, Y, and Ba. We also measured $EW$s by hand for two special and delicate cases, namely oxygen and lithium using the {\scshape iraf} {\scshape splot} task.
For lithium, we employed a classical analysis using the 6707.8\,\AA{} line, taking into account the nearby \ion{Fe}{i} line at 6707.4\,\AA{}. We measured the $EW$ by hand (in m\AA{}), using the 
{\scshape iraf} {\scshape splot} task. We use the standard notation, where $A($Li$)= \log\,$[n(Li)/n(H)]\,+\,12. In the case of oxygen, we used the [\ion{O}{i}] 6300\,\AA{} line. This oxygen line is blended with a \ion{Ni}{i} feature; we corrected the $EW$s 
accordingly by using the  methodology and line atomic parameters described in \citet{bel15}. Finally we also derive rubidium abundances using stellar synthesis for the 
7800\,\AA{} \ion{Rb}{i} line, following the methodology in \citet{dor13}.

Table~\ref{Abund} summarises the abundances found for each star together with their errors. We estimate the cluster average by using a weighted arithmetic mean (employing the variances as weights). We computed conservative errors, since the 
associated uncertainty is the combination in quadrature of the individual typical error and the star-to-star dispersion. This average has been computed both with and without star 24, since it has a blue companion. Although we do not 
see any trace of the companion in the spectral range used for the analysis, we cannot discard that it is affecting the results, given the values found are quite different from 
those of stars 9 and 66, which have very similar spectra.

\begin{table*}
\caption{Chemical abundances, relative to solar abundances by \citet{Gre07}, measured first on the cool stars and then only with stars 9 and 66, since star 24 might be polluted by its blue companion. }
\begin{center}
\begin{tabular}{lcccccc}   
\hline\hline
Star & [O/H] &[Na/H] & [Mg/H] & [Si/H] & [Ca/H] & [Ti/H]\\
\hline
9  & $-$0.24 $\pm$ 0.26 & $-$0.04 $\pm$ 0.35 & $-$0.01 $\pm$ 0.15 &    0.11 $\pm$ 0.32 & 0.07 $\pm$ 0.26 & 0.21 $\pm$ 0.32\\
24 &    0.26 $\pm$ 0.17 &    0.02 $\pm$ 0.11 &    0.02 $\pm$ 0.11 &    0.15 $\pm$ 0.23 & 0.38 $\pm$ 0.20 & 0.27 $\pm$ 0.16\\
66 & $-$0.48 $\pm$ 0.27 & $-$0.07 $\pm$ 0.45 & $-$0.12 $\pm$ 0.14 & $-$0.06 $\pm$ 0.36 & 0.24 $\pm$ 0.39 & 0.07 $\pm$ 0.45\\

\hline
Mean & $-$0.02 $\pm$ 0.44  & 0.01 $\pm$ 0.31 & $-$0.03 $\pm$ 0.15 &    0.09 $\pm$ 0.32 & 0.26 $\pm$ 0.32 & 0.24 $\pm$ 0.33\\
Mean (no 24) & $-$0.36 $\pm$ 0.31  & $-$0.05 $\pm$ 0.40 & $-$0.07 $\pm$ 0.16 &    0.03 $\pm$ 0.36 & 0.12 $\pm$ 0.35 & 0.16 $\pm$ 0.40\\
     &                     &                    &                    &                    &                    & \\

\hline\hline
Star & [Ni/H] & [Rb/H] & [Y/H] & [Ba/H] & $EW$(Li) & A(Li)\\
\hline

9  &    0.00 $\pm$ 0.34 & $-$0.31            & $-$0.28 $\pm$ 0.30 & 0.48 $\pm$ 0.16 & 164.2 & $<$\,0.89 \\
24 &    0.25 $\pm$ 0.20 &    0.05            &    0.09 $\pm$ 0.40 & 0.58 $\pm$ 0.11 &  99.0 & $<$\,0.93 \\
66 & $-$0.15 $\pm$ 0.37 & $-$0.35            & $-$0.33 $\pm$ 0.52 & 0.35 $\pm$ 0.19 & 122.9 & $<$\,0.52 \\

\hline
Mean &  0.13 $\pm$ 0.36 & $-$0.20 $\pm$ 0.22 & $-$0.18 $\pm$ 0.47 & 0.51 $\pm$ 0.19 & --- & ---\\
Mean (no 24) &  $-$0.07 $\pm$ 0.37 & $-$0.33 $\pm$ 0.03 & $-$0.29 $\pm$ 0.41 & 0.43 $\pm$ 0.20 & --- & ---\\
\hline

\end{tabular}

\label{Abund}
\end{center}
\end{table*}


\section{Discussion}

\subsection{Cluster parameters}

Unlike previous works, it should be noted that our photometric analysis has been strengthened by including spectroscopy. In addition, we have employed the largest number of likely members to date.
In Table~\ref{cum_param}, we summarise our values for cluster parameters and compare them with those in the literature. 
The amount of reddening is compatible within the errors with the values calculated by
previous authors, with the exception of \citet{Pau05}, who used the $\Delta\,a$ photometric system instead of Johnson's. Our determination of a low metallicity for the cluster implies that it is slightly older 
and is located somewhat closer than in previous works, which assumed solar composition. With respect to the first studies, based on photographic and photoelectric data, we observe a larger cluster size. 
Our result is consistent with the membership of S66, a radial velocity member, that is located at around $2\farcm2$ from the cluster centre. Morever, we provide a new value for the cluster coordinates (10h\:00m\:41s, $-54\degr47\arcmin21\arcsec$) 
very close to the nominal ones, with a small offset (this work\,$-$\,nominal) of $\Delta\alpha\,\approx\,2\arcsec$ and $\Delta\delta\,\approx\,3\arcsec$.

\begin{table*}
\caption{Summary of cluster parameters for NGC~3105 derived in this work compared to studies found in the literature.} 
\begin{center}
\begin{tabular}{lccccc}   
\hline\hline
Reference    & Phot. &  $E(B-V)$     & $d$ (kpc)       &  Age (Ma)  & $r_{\textrm{cl}}$ (pc)\\
\hline     
\citet{Mo74} & Pg+Pe & 1.09\,$\pm$\,0.03 & 8.0\,$\pm$\,1.5  & ---           & 2.6            \\ 
\citet{Fi77} & Pe    & 1.08\,$\pm$\,0.03 & 5.5\,$\pm$\,0.8  & ---           & 1.8            \\
\citet{Sa01} & CCD   & 1.06              & 9.5\,$\pm$\,1.5  & 25\,$\pm$\,10 & ---            \\
\citet{Pau05}& CCD   & 0.95\,$\pm$\,0.02 & 8.5\,$\pm$\,1.0  & 20\,$\pm$\,5  & ---            \\
This work    & CCD   & 1.03\,$\pm$\,0.03 & 7.2\,$\pm$\,0.7  & 28\,$\pm$\,6  & 5.8\,$\pm$\,1.7\\
\hline
\end{tabular}

\label{cum_param}
\end{center}
\end{table*}

NGC~3105, located at $R_{\textrm{GC}}\sim$10\,kpc and presenting a [Fe/H]$\sim-$0.3, is a distant and metal-poor cluster.
It is located at a position compatible with that expected from its observed radial velocity, computed by us for the first time. According to the flat rotation 
curve for the Galaxy \citep{reid09}, in the line-of-sight direction to the cluster, the expected distance corresponding to the observed RV is 7.0\,kpc, compatible, within the errors, with the value found in this work, $d=7.2\pm0.7$\,kpc.
In addition, we estimated, also for the first time, the initial mass for this cluster obtaining a value around 4\,000\,M$_{\sun}$, which makes NGC~3105 a moderately massive cluster.

Given its location in the Galaxy ($l\sim280\degr$, $d\sim$7\,kpc) NGC~3105 is young enough to serve as a tracer of the Sagittarius-Carina arm.
However, in this direction, and at similar distances, there are no clusters with known metallicity with which we can compare our results.

In NGC~3105 we find RSGs with masses around 9.5\,M$_{\sun}$. According to most modern evolutionary tracks, stars of this mass at this metallicity, will end their lives in 
a core-collapse supernova explosion, but are just above the minimum mass required for this to happen \citep[see][and references therein]{Doherty17}. Star 33, for which unfortunately we lack 
a high-resolution spectrum, is definitely an M-type supergiant, while stars 9, 24 and 66 have K types. In a sample of clusters with ages between 50 and 100 Ma, all 
the evolved red stars (which will become AGB stars according to models) have K types \citetext{\citealp{Roma}; see also \citealp{Be55}}. In clusters somewhat younger than NGC~3105, such as NGC~3766 or NGC~663, 
there are only M-type supergiants. This is in agreement with the results of \citet{Me812} who found that the average spectral type of red (super)giants becomes later 
with younger ages. The comparison, however, is not straightforward, because the typical spectral types of RSGs depend on metallicity \citep{Levesque06,ric16}. In fact, the spectral distribution of the 
RSGs in NGC~3105 is similar to that of moderate-to-low-luminosity RSGs in the LMC, which presents a similar average metallicity 
\citep{ric16}. It seems thus that the stars in NGC~3105 are among the lowest-mass stars that will give rise to a core-collapse supernova. 

\subsection{Blue to red supergiant ratio}

The number of blue supergiants per red supergiants (B/R ratio) is one of the most important problems in high-mass stellar evolution. Current models fail when trying to reproduce the observed ratios at different
metallicities \citep{lan95,dohm02}. This ratio is an excellent test for stellar models because it is very sensitive to the input physics: mass loss, convection and mixing process \citep{lan95} or rotation \citep{mae01}.
The B/R ratio seems to be directly proportional to metallicity \citep{lan95,egg02}, with consequences such as the nature of the supernova progenitors in different environments
\citep{lan91}, the study of the luminous star population in other galaxies \citep{lan95,ori99} or the variation in the period of variable stars \citep{mae01}.

According to \citet{egg02}, who studied this ratio in young clusters at different metallicities in the Milky Way (MW) and the
Magellanic Clouds (SMC and LMC), the B/R ratio as a function of metallicity is given by this expression: 
\begin{equation}
  \frac{B/R}{(B/R)_{\sun}}\,=\,0.05\,\cdot\,\mathrm{e}^{3\,\cdot\,\frac{Z}{Z_{\sun}}}  
\end{equation}
where B is the number of BSGs (including O, B and A supergiants), R is the number of RSGs (K and M), (B/R)$_{\sun}\cong3$ and Z$_{\sun}$=0.02. In our case, for NGC~3105 (not included in the work of \citealt{egg02}), 
$Z\cong0.0103$ (referred to the solar value just mentioned, Z$_{\sun}$=0.02) and so we would expect an ``empirical'' ratio (B/R)$_{\textrm{emp}}\approx0.70$. Since the cluster hosts 2\,A supergiants and 5 RSGs (3\,K + 2\,M), the ratio observed is (B/R)$_{\textrm{obs}}=0.4$. This 
value is slightly more than half of the expected empirical value, though we caution about the low statistical significance of our sample.

Another interesting observable to constrain theroretical models is the yellow-to-red supergiant (Y/R) ratio. This ratio is very sensitive to the extension of 
the blue loop \citep[where YSGs spend most of the time during this phase, according to][]{An14}. The blue loop depends strongly on the metallicity and the input physics \citep{mat82,eks12,wal15}.
NGC~3105 contains one YSG and five RSGs, i.e. (Y/R)=0.2. This ratio is typically found in clusters within a similar age range as NGC~457, NGC~654, NGC~6649, NGC~6664 or Berkeley 55 \citep[][and references therein]{YSG_OC,Be55}.
However, this value is smaller than those predicted by stellar evolution models.

\subsection{Be stars}

Classical Be stars are non-supergiant B-type stars that show or have shown Balmer lines in emission at any time \citep{Po03}. This phenomenon is related to fast rotation, which generates circumstellar discs around the star, 
responsible for the emission observed. Be stars are usually located around the MSTO of young open clusters \citep{McSw05}. They are observed in clusters covering the whole B-type range. \citet{Me82} suggested that the Be fraction, i.e. Be/(B+Be),
decreases with increasing age. He found a Be star distribution with two peaks: one at B1\,--\,B2 and the other at B7\,--\,B8. According to theoretical considerations concerning rotational velocity distribution \citep{Gra16}, the largest fraction
of Be stars should occur in populations ranging from $\approx$16 to $40\:$Ma. \citet{Gra16} assume that the Be phenomenon is observed above $\Omega/\Omega_{\textrm{crit}}\geq0.8$. In addition, the Be fraction is around
three times higher in early B stars. These results seem to agree with observations: NGC~663 \citep{Pi01} or NGC~7419 \citep{7419}, with ages in the 15\,--\,25 Ma range, are among the clusters with the highest Be fractions (up to 
$\approx$40 per cent). On the other hand, \citet{McSw05} performed a detailed analysis of a sample containing 55 open clusters, finding a broad maximum for the Be fraction for ages between 25\,--\,100~Ma. 

In the field of NGC~3105 we find 11 Be stars, from which we consider nine as likely members. Most of them are located in the upper main sequence, close to the MSTO. We find seven Be stars between the top of the MS and spectral type b3.
This number implies a Be fraction of 35\,$\%$ for stars brighter than $V=16.3$, corresponding to spectral type B3\,V, or 25\,$\%$ when considering likely members with spectral types down to b3. 
\citet{tar12} studied the relative content of Be stars in 42 young clusters aged up to 35 Ma, taking 
into account objects in the B0\,--\,B3 range only. The maximum frequency of Be stars occur in clusters with ages from 13 to 25 Ma, reaching a Be fraction around 25\,$\%$. For the age range of NGC~3105 (in their figure 2, the 
25\,--\,30 Ma bin) they found a fraction 15$\%$, almost half that found by us in NGC~3105. Hence, NGC~3105 (not included in their work) is among the clusters with the highest Be fractions 
together with well-known clusters such as h/$\chi$~Persei \citep{Malchenko08} or NGC~7419 \citep{7419}. However, these clusters are younger than NGC~3105, with ages around 12--14 Ma.     

The shape of the cluster can be taken as a clue to the initial angular momentum with which the cluster originated. According to \citet{kel99} the ellipsoidal shape of the SMC massive cluster NGC~330 would be the consequence
of having been formed from material with high angular momentum. This would explain the high Be fraction observed in that cluster, related to a rapid rotation rate. This same argument might be used as an explanation for the high Be frequency in NGC~3105, a cluster with a high eccentricity ($\epsilon=0.51$, see Sect.~\ref{sec_geom_3105}).
However, we are cautious since this ellipsoidal shape could also be the result of a previous merger of two sub-clusters.

Finally, the Be phenomenon is also related to metallicity. In fact, the Be fraction increases when decreasing metallicity \citep{mae00,Mar07}. According to \citet{mae00}, in low metallicity environments, mass loss via a radiative wind is reduced and thus not very effective at breaking down fast rotation. \citet{Mar07} studied the ratios $\Omega/\Omega_{\textrm{crit}}$ for Be stars in several galaxies with different metallicities. They found for the MW (assuming $Z=0.020$),
LMC ($Z=0.004$) and SMC ($Z=0.002$) $\Omega/\Omega_{\textrm{crit}}$ ratios of 0.80, 0.85 and 0.95, respectively. In these galaxies they also studied the B and Be populations and noted for field Be stars fractions of 15, 17.5 
and 26 per cent. As seen, the frequency of Be stars is inversely related to the average metallicity. NGC~3105 has a particularly low metallicity for a Milky Way young cluster, and seems to fit this trend.

\subsection{Low metallicity}\label{sec:low_met}

\begin{figure}  
  \centering         
  \includegraphics[width=\columnwidth]{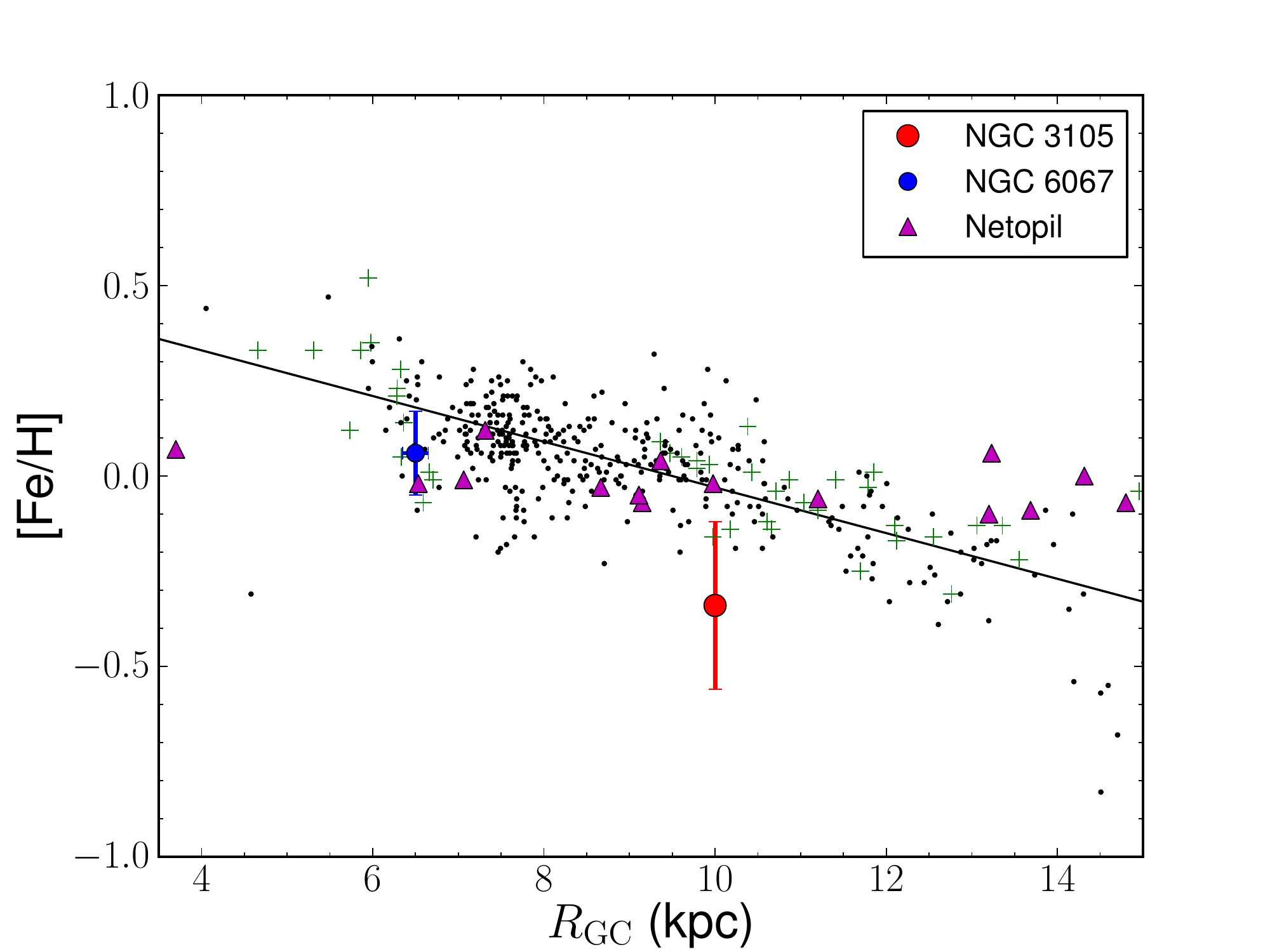}   
  \caption{Iron abundance gradient in the Milky Way found by \citet{Ge13,Ge14}. The black line is the Galactic gradient ($-$0.06~dex/kpc), green crosses are Cepheids studied in those papers,
  whereas black dots show data for other Cepheids from literature used by these authors. Magenta triangles represent the young open clusters ($<500\:$Ma) in the sample compiled by \citet{net16}. Finally, the 
  red circle is NGC~3105 and the blue one is NGC~6067. Both young clusters have been analysed by our group with the same techniques. NGC~3105 falls well below the metallicity typically observed at its Galactocentric distance. All the 
  values shown in this plot are rescaled to \citet{Ge14}, i.e. R$_{\sun}$=7.95 kpc and A(Fe)=7.50.} 
  \label{gradient}  
\end{figure}

In the current work, for the first time, chemical abundances for NGC~3105 are provided. In the absence of previous work with which to compare our results, we place the cluster on the Galactic gradient. As a reference we use the
work by \citet{Ge13,Ge14}. They used Cepheids for estimating the radial distribution of metallicity (in terms of [Fe/H]), as this is a young population ($\tau\approx20$\,--\,400~Ma) tracing present-day abundances, and thus providing a good comparison for young clusters. In Fig.~\ref{gradient}, we show the Galactic gradient 
by \citet{Ge13,Ge14} with the position of NGC~3105 overplotted as well as a sample of open clusters taken from \citet{net16} for further comparison. We selected young clusters ($\tau<500\:$Ma) whose 
metallicity is obtained from high quality spectra \citep{hei14} as the average of, at least, three stars. It is obvious that NGC~3105 is a metal-poor cluster. 

\begin{figure*}  
  \centering         
  \includegraphics[width=19cm]{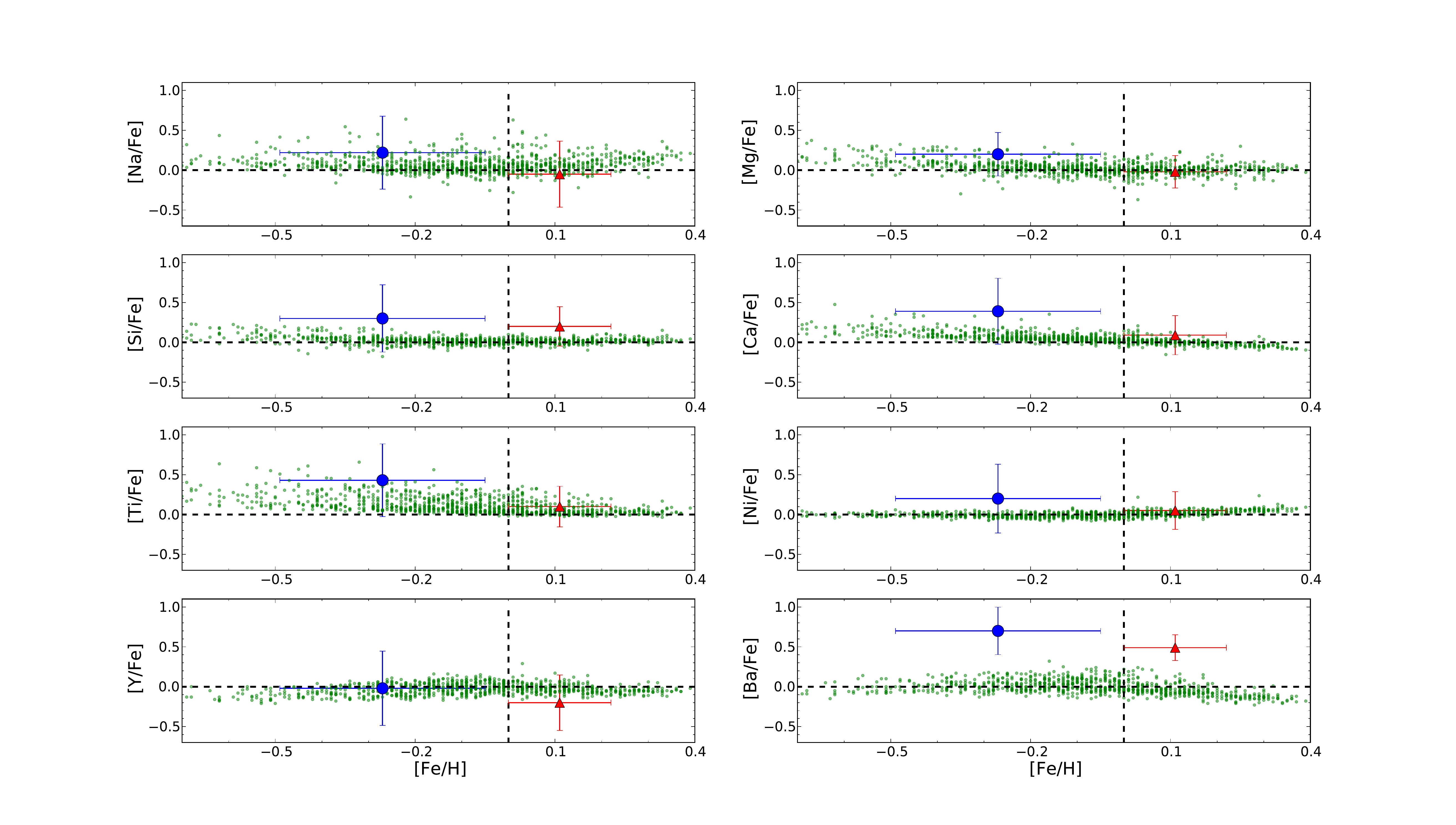}   
  \caption{Abundance ratios [X/Fe] versus [Fe/H]. The green dots represent the galactic trends for the thin disc \citep{vardan,elisa17}. NGC~6067 is the red triangle whereas 
  NGC\,3105 (without including S24) is the blue circle. Clusters are represented by their 
  mean values. The dashed lines indicates the solar value.} 
  \label{trends}  
\end{figure*}

We also compare our abundances with the Galactic trends for the thin disc (Fig.~\ref{trends}). We plot abundance ratios [X/Fe] vs [Fe/H] obtained by \citet{vardan} for Na, Mg, Si, Ca, Ti and Ni and by \citet{elisa17} for Y and Ba in the workframe 
of the HARPS GTO planet search program. The chemical composition of NGC\,3105 is compatible, within the errors, with the Galactic trends observed in the thin disc of the Galaxy.
No significant amounts of either Li or Rb have been found in the RSGs.
We derived a roughly solar [Y/Fe] against a supersolar [Ba/Fe], which is in good agreement with the dependence on age and Galactic location found by \citet{Mi13} by comparing the
abundances of Y and Ba in different open clusters. 
Remarkably, we find a strong over-abundance of Barium ([Ba/Fe]=$+$0.8). Several young open clusters have been observed to have Ba abundances higher than those predicted by standard theoretical models.
To explain this enrichment of Ba, \citet{dor09} suggested an enhanced ``$s$-process''. Our result supports this idea (see their figure~2).

\subsection{The nature of star 406}\label{406}

Star 406 has been found to be a luminous late-M star that we have classified as M6\,Ib, even though luminosity criteria are not well defined at these spectral types \citep[cf.][]{neg12}. 
The star has a radial velocity fully compatible with cluster membership, and has thus been included as a cluster member. However, we must note that we know of no other star with a similar spectral type that is a member of an open cluster of this age. Two massive young clusters, Stephenson~2 (16\,--\,$20\:$Ma) 
and NGC~7419 ($\approx14\:$Ma), contain stars with spectral types M6\,--\,7 that are clearly supergiants. These objects, however, are much brighter than other RSGs in those clusters, and show evidence for very heavy mass loss. Star 406 is fainter in $K_{\textrm{S}}$ than the other M-type supergiant, S33. Using 
the ($K_{\textrm{S}}-W3$) colour as an indication of mass loss, it does not seem to have a significantly higher mass loss than the other RSGs (0.38~mag against 0.30~mag for S33 or 0.33 for S66). However, 
interpretation of the data is difficult, as S33 is almost 
one magnitude brighter than the next brightest RSG, S66. The possibility that S406 is an unrelated luminous red giant with exactly the same radial velocity falling within the tidal radius of the cluster is slim.
Given the masses of the evolved stars, close to the boundary between intermediate- and high-mass stars, we could speculate that S406 might be a super-AGB star (i.e. a star of $\ga8\:$M$_{\sun}$ that can ignite Carbon off-center but will still experience a third dredge-up; \citealt{Po08}). 
Against this interpretation, we note that S406 does not show any evidence for the presence 
of Li or Rb, which are seen in many massive AGB stars at LMC typical metallicity and similar temperatures \citep{anibal09}.

\section{Conclusions}

We have performed the most complete analysis to date of NGC~3105, a cluster poorly studied until now. 
Our photometry is deeper than previous studies and no significant offsets with them are noted. The cluster parameters obtained in this work --\,$E(B-V)=1.03$, $d=7.2$~kpc, $\tau$=27\,Ma\,-- are compatible, within
the errors, with those found in the literature. 

We estimate a cluster size, $r_{\textrm{cl}}\approx$2.7 arcmin, doubling or even
tripling previous estimations. This size is consistent with the location of the most distant members confirmed via radial velocity. We identify 126 likely (blue) members from which we derive
an initial mass slightly above $4\,000\:$M$_{\sun}$. Therefore, NGC~3105 is a moderately massive cluster, as expected from a population of eight supergiant stars, spanning the blue, yellow and red phase, a rare ocurrance. 
Thus, NGC~3105 is comparable to NGC~7419 as a template for studying obscured RSGCs and for constraining
evolutionary models of intermediate- and high-mass stars. As in NGC~7419, we find a cluster with less than $10\,000\:$M$_{\sun}$ (in this cases, significantly less), that still contains five RSGs (one is not a certain member).

Evolved stars in NGC~3105 are set in the lower limit of massive stars. Its RSGs, with masses around 9.5\,M$_{\sun}$, are just above the minimum mass to explode as a SN. 

As in other clusters, we find that the location of the A-type supergiants in both the colour-magnitude and Kiel diagrams indicate that they are just leaving the MS 
and are not blue-loop objects. Current Geneva models \citep{georgy13} suggest that most A-type supergiants are He-core burning objects in the blue loop, but there is little 
observational evidence for this. In fact, even the F-type supergiant seems to be in the Hertzsprung gap, where no stars are expected, rather than in the blue loop.

Based on good $S/N$ and high-resolution spectra, we infer a cluster $v_{\textrm{rad}}$\,=\,$+$46.9\,km\,s$^{-1}$ and  derive a subsolar metallicity, [Fe/H]=$-$0.29. At the cluster Galactocentric distance ($R_{\textrm{GC}}$=10.0 kpc) this value is rather below
the expected value according to the Galactic gradient. This low metallicity, as well as its elliptical geometry would explain the high fraction ($\approx$25 per cent) of Be stars found from the stars observed with the earliest spectral types down to
spectral type b3. 

For the first time, we have calculated stellar parameters for hot and cool stars within the cluster. As shown in the Kiel diagram, the parameters obtained for both 
sets of stars, using different codes and techniques, agree well between them (as they trace a single theoretical isochrone) and also with the results derived from 
photometry (as it is the isochrone corresponding to the photometric age).
For the first time we also have determined chemical abundances of Li, O, Na, $\alpha$-elements (Mg, Si, Ca and Ti), Fe-group (Ni) and $s$-elements (Rb, Y and Ba). The over-abundance of Ba found
supports the ``$s$-enhanced'' process. 

Determination of the metallicity of a young cluster is fundamental for its correct characterisation. The use of solar metallicity isochrones leads to a an incorrect 
estimation of the age and the mass of the RSGs. Despite its unusually low metallicity, NGC~3105 is an excellent laboratory to improve theoretical models. Derivation of stellar 
parameters for its blue supergiants and its brightest RSG, star 33, would provide an excellent complement to this work and improve the value of NGC~3105 as a template for 
more obscured or unresolved clusters.

\section*{Acknowledgements}

We thank the anonymous referee for his/her helpful suggestions which have helped to improve this paper.
We also thank An\'{\i}bal Garc\'{\i}a Hern\'andez for instruction in the use of {\scshape turbospectrum}. 
This research is partially supported by the Spanish Government Ministerio de Econom\'{i}a y Competitividad under grants BES 2013-065384 and AYA2015-68012-C2-2-P (MINECO/FEDER).
H.M.T. would like to thank support from the Spanish Government (Ministerio de Econom\'{i}a y Competitividad) under grant FJCI-2014-23001.
The research leading to these results has received funding from the European Community's Seventh Framework Programme (FP7/2013-2016) under grant agreement number 312430 (OPTICON); 
proposal number 2015A/008.
This paper uses observations made at the South African Astronomical Observatory (SAAO).
This research has made use of the Simbad database, operated at CDS, Strasbourg (France). This publication also made use of data products from the Two Micron All Sky Survey, which is a joint project of the 
University of Massachusetts and the Infrared Processing and Analysis Center/California Institute of Technology, funded by the National Aeronautics and Space Administration and 
the National Science Foundation. 




\bibliographystyle{aa}
\bibliography{latex} 

\appendix
\section{Additional tables}
\begin{landscape}
\begin{table}
\caption{$UBVR$ photometry obtained in this work for 607 stars in the field of NGC 3105. Coordinates and, when possible, crossmatched ID with 2MASS are also shown. Only ten rows are listed here for guidance regarding its 
form and content. Full table is available online.}
\begin{center}
\begin{tabular}{lcccccccccccc}   
\hline\hline
ID  &  RA(J2000) & DEC(J2000) & $V$ & $\sigma_V$ & $(B-V)$ & $\sigma_{(B-V)}$ & $(U-B)$ & $\sigma_{(U-B)}$ & $(V-R)$ & $\sigma_{(V-R)}$ & $N$ & 2MASS\\
\hline

  1     & 150.116411024   & -54.768803513 & 18.065   & 0.062	& 1.172    & 0.033    & 0.518	  & 0.048   & 0.692   & 0.086	& 2	 & 10002796-5446081\\
  2     & 150.116857743   & -54.798684069 & 17.864   & 0.005	& 1.831    & 0.007    & 1.406	  & 0.024   & 1.020   & 0.010	& 1	 & 10002807-5447552\\
  3     & 150.116930869   & -54.806688297 & 19.449   & 0.015	& 1.366    & 0.018    & 0.997	  & 0.053   & 0.722   & 0.030	& 1	 & 10002807-5448243\\
  4     & 150.117564385   & -54.784364293 & 16.503   & 0.046	& 0.913    & 0.019    & 0.292	  & 0.011   & 0.580   & 0.057	& 3	 & 10002823-5447038\\
  5     & 150.117598652   & -54.791542491 & 18.775   & 0.048	& 0.932    & 0.019    & 0.689	  & 0.057   & 0.529   & 0.081	& 2	 & \\		 
  6     & 150.118010282   & -54.757280988 & 19.930   & 0.019	& 1.361    & 0.024    & 0.583	  & 0.054   & 0.739   & 0.035	& 1	 & \\		 
  7     & 150.118054450   & -54.763392111 & 19.416   & 0.016	& 1.302    & 0.019    & 0.800	  & 0.077   & 0.709   & 0.028	& 1	 & 10002833-5445487\\
  8     & 150.118559593   & -54.799973416 & 18.988   & 0.012	& 1.455    & 0.014    & 0.984	  & 0.038   & 0.799   & 0.019	& 1	 & 10002848-5448001\\
  9     & 150.118791460   & -54.807744006 & 17.004   & 0.039	& 0.884    & 0.040    & 0.230	  & 0.051   & 0.515   & 0.058	& 6	 & 10002855-5448278\\
  10    & 150.118934401   & -54.810798820 & 17.796   & 0.055	& 1.297    & 0.083    & 1.039	  & 0.057   & 0.667   & 0.089	& 3	 & 10002861-5448386\\

\hline
\end{tabular}

\label{phot}
\end{center}
\end{table}

\begin{table}
\caption{$Q$ index, photometric spectral types and reddening for the likely blue members of NGC~3105. Only ten rows are listed here for guidance 
regarding its form and content. Full table is available online.}
\begin{center}
\begin{tabular}{lccc}   
\hline\hline
ID  &  $Q$ & SpT & $E(B-V)$\\
\hline
4 & -0.40 & b5 & 1.14\\ 
9 & -0.44 & b5 & 1.12\\ 
19 & -0.10 & b9 & 1.15\\ 
24 & -0.13 & b9 & 1.17\\ 
33 & -0.20 & b8 & 1.15\\ 
45 & -0.38 & b7 & 1.05\\ 
54 & -0.57 & b3 & 1.10\\ 
65 & -0.43 & b5 & 1.21\\ 
78 & -0.08 & b9 & 1.13\\ 
80 & -0.25 & b8 & 1.03\\ 
\hline
\end{tabular}
\label{phot_spt}
\end{center}
\end{table}
\end{landscape}

\end{document}